\documentclass{aa}  

\newcommand{\cii}{[C\,{\sc ii}] }
\usepackage{graphicx}
\usepackage{natbib}
\usepackage[breaklinks=true]{hyperref}
\hypersetup{colorlinks=true,linkcolor=red,citecolor=blue,filecolor=cyan,urlcolor=magenta}
\bibpunct{(}{)}{;}{a}{}{,} 
\usepackage{graphicx}
\usepackage{txfonts}

\begin{document} 

   \title{Resolving the molecular gas emission of the $z\sim2.5-2.8$ starburst galaxies SPT\,0125$-$47 and SPT\,2134$-$50}

   \author{K. Kade
          \inst{1}
          \and
          M. Bredberg \inst{2}
          \and
          K. Knudsen \inst{1}
          \and
          S. K\"onig \inst{1}
          \and
          G. Drouart \inst{1}
          \and
          A. B. Romeo \inst{1}
          \and
          T.J.L.C. Bakx \inst{1}  
          }

   \institute{Department of Space, Earth \& Environment, Chalmers University of Technology, SE-412 96
              Gothenburg,  Sweden  \\ \email{kiana.kade@chalmers.se}
              \and
              Institute of Physics, Laboratory of Astrophysics, École Polytechnique Fédérale de Lausanne (EPFL), Observatoire de Sauverny, Versoix, 1290, Switzerland
             }

   \date{Received July 11, 2024; accepted November 11, 2025}
 
  \abstract
   {The comoving cosmic star formation rate density peaks at $z \sim 2-3$, with dusty star-forming galaxies being significant contributors to this peak. These galaxies are characterized by their high star formation rates and substantial infrared (IR) luminosities. The formation mechanisms remain an open question for these galaxies, particularly with respect to how such intense levels of star formation are triggered and maintained.}
   {We aim to resolve CO(3--2) emission toward two strongly lensed galaxies, SPT\,0125-47 and SPT\,2134-50, at $z \sim 2.5-2.8$ to determine their morphology and physical properties.}
   {We used high-resolution ALMA band 3 observations of CO(3--2) emission toward both sources to investigate their properties. We performed parametric and nonparametric lens modeling using the publicly available lens modeling software {\sc PyAutoLens}. We divided the CO(3--2) emission line into two bins corresponding to the red and blue portions of the emission line and nonparametrically modeled the source plane emission for both bins.}
   {We found that both sources are well described by a single S\'ersic profile in both the parametric and nonparametric models of the source plane emission, in contrast to what was previously found for SPT\,0125-47. Parametric lens modeling studies of the red and blue bins have reported distinctive differential magnification across the line spectrum. We performed a basic analysis of the morphology and kinematics in the source plane using nonparametric lens modeling of the red and blue bins. We found tentative evidence of a velocity gradient across both sources and no evidence of any clumpy structure, companions, or ongoing mergers.}
   {The previously calculated high star formation rates and low depletion times of both SPT\,0125-47 and SPT\,2134-50 suggest that these galaxies are undergoing a dramatic phase in their evolution. Given the lack of evidence of ongoing interactions or mergers in our source plane models, we suggest that the intense star formation was triggered by a recent interaction and/or merger. We also consider the possibility that these galaxies might be in the process of settling into disks.}


   \maketitle

\section{Introduction}

The comoving cosmic star formation rate density is known to peak between $z\sim 2-3$ \citep{Madau14}. A primary contributor to this peak is dusty star forming galaxies \citep[DSFGs;][]{Smail98, Blain02, Casey14}\footnote{These galaxies are often also referred to as submillimeter galaxies (SMG). Historically, SMGs have been defined as being bright at submillimeter wavelengths, typically with 850\,$\mu$m flux density $\mathrm{S_{850}} > 3-5$\,mJy. Both galaxies in this work would be SMGs in this classification methodology.}. These galaxies, occurring at redshifts $z > 2$, typically do not appear in optical wavelengths due to the UV light from their intense starbursts being reprocessed by dust into infrared (IR) wavelengths \citep[e.g.,][]{Smail97, Hughes98, Bertoldi00}. These galaxies are known to have prodigious star formation rates of $\geq 1000$\,M$_{\odot}$\,yr$^{-1}$ and high IR luminosities of $\sim10^{13}$\,L$_{\odot}$ \citep[e.g.,][]{Barger14, Swinbank14, Simpson14}. They have also been found to have relatively large molecular gas masses of $10^{10} - 10^{11}$\,M$_{\odot}$ and, thus, low molecular gas depletion timescales of $\lesssim 1$\,Gyr \citep[e.g.,][]{Bothwell13, Aravena16, Birkin21}. DSFGs are considered to be similar to local ultraluminous IR galaxies \citep[ULIRGs;][]{Sanders96}. Both have properties in common, including high molecular gas masses and IR luminosities \citep[e.g.,][]{Tacconi08, Engel10, Riechers11, Bothwell13}. Given these properties, it has been suggested that these galaxies will evolve into massive early-type galaxies \citep[e.g.,][]{Simpson14, Birkin21}.

\begin{table*}[t]
    \centering
    \caption{Details of the observations.}
    \begin{tabular}{l c c c c c c} \hline \hline
        Source & Emission line$^a$ & Date of Obs.$^b$ & $\nu_{\mathrm{spw, central}}^c$ & Native channel width$^d$ & Synthesized beam$^e$ & RMS$^f$\\
         & & [yyyy mm dd] & [GHz] & [MHz] & [$'' \times '', ^{\circ}$] & [mJy/beam]\\ \hline 
        
        SPT\,0125-47 & CO(3--2) & 2017 09 21 & 98.476 & 3.9 & $0.11 \times 0.11, 33$ & 0.4 \\
        SPT\,2134-50 & CO(3--2) & 2017 09 25 & 91.495 & 3.9 & $0.15 \times 0.13, 76$ &  0.35 \\ \hline
        
    \end{tabular}
    \tablefoot{
    \tablefoottext{a}{Observed emission line.}
    \tablefoottext{b}{Date of ALMA observation.}
    \tablefoottext{c}{Central frequency of the spectral window containing the specific line.}
    \tablefoottext{d}{Native channel width of the calibrated and imaged data for the specific line.}
    \tablefoottext{e}{Synthesized beam of the spectral window using natural weighting and a spectral resolution of \(\sim70\,\mathrm{km\,s}^{-1}\).}
    \tablefoottext{f}{Per-channel root mean square (RMS) of the spectral window containing the CO(3--2) emission line using natural weighting and a spectral resolution of \(\sim70\,\mathrm{km\,s}^{-1}\).}}
    \label{tab:obs_details}
\end{table*}

One primary open question is how these galaxies obtain the fuel to maintain such intense star formation rates (SFRs). One explanation for this is based on the occurrence of mergers. Indeed, DSFGs have been shown to represent overdensities and, in some cases, show clear signs of mergers or interactions \citep[e.g.,][]{Brodwin08, Viero09, Daddi09, Capak11, Kade23}. This is also consistent with simulations of massive galaxy evolution which have indicated the importance of mergers at high-redshift \citep[e.g.,][]{Hopkins08}. Other studies have shown DSFGs can be single disk-like galaxies \citep[e.g.,][]{Hodge19, Rizzo21, Amvrosiadis25}. Alternatively, the high SFRs observed in these galaxies may result from intensely star-forming clumps. Studies searched for these clumps in IR wavelengths with mixed results \citep[e.g.,][]{Swinbank10, Swinbank15, Iono16, Oteo17}. \citet{Hodge16} found no evidence of clumps in dust continuum emission of 16 DSFGs, while \citet{Spilker22} found clear evidence of clump-like structures in the $z = 6.9$ DSFG SPT\,0311–58. Simulations suggest these clumps may occur on scales of $200-500$\,pc \citep[e.g.,][]{Dekel09, Bournaud14}, in good agreement with studies that detect these clumps. Observations of these clumps in high-redshift galaxies can provide additional constraints on mechanisms for star formation and the relation between stability and turbulence \citep[][]{Romeo10, Romeo14}. However, the angular resolution necessary to resolve these scales is observationally challenging and expensive. 

One method to improve the angular resolution of observations is to observe gravitationally lensed galaxies. In galaxy-galaxy lensing scenarios, this phenomenon can effectively improve the angular resolution of the observations by $\sqrt{\mu}^{-1}$, assuming a constant magnification factor across the image. Thus, high angular resolution observations from, for example, the Atacama Large Millimeter/Submillimeter Array (ALMA) can resolve down to the scales necessary to determine the morphology of DSFGs and determine whether they are composed of clumps or interacting galaxies. Studies investigating the properties of lensed DSFGs have become an increasingly common method for investigating the nature of these galaxies \citep[e.g.,][]{Spilker22, Amvrosiadis25}.

Indeed, studies often use observations of higher-$J$ carbon monoxide (CO) transitions or brighter far-IR (FIR) emission lines such as \cii fine structure emission line. However, lower-$J$ CO transitions trace more diffuse molecular gas and are therefore key to determining the stability of clumps and disks. \citet{Canameras17} used CO(4--3) emission to study the strongly lensed $z = 3.0$ starburst galaxy, PLCK G244.8+54.9, with a source plane resolution of down to 60\,pc. The authors found that at the given resolution, they were able to both observe clumpy structure and perform an investigation into the stability of the disk, finding that the disk was, in fact, stable while still hosting a massive starburst \citep{Canameras17}. In general, however, observations of lower-$J$ CO transitions are seldom performed due to their observationally costly nature.

This paper studies the two strongly lensed DSFGs SPT\,0125-47 and SPT\,2134-50, originally discovered as bright sources in South Pole Telescope (SPT) data. These sources were originally selected to have $\rm S_{1.4 mm} > 20$\,mJy, exhibit a dusty spectrum, and have no bright radio or far-IR (FIR) counterparts in the SPT data \citep{Weiss13}. The original redshift detections for these sources come from a blind survey of CO(3--2) emission from \citet{Weiss13}. In addition, dust continuum emission has been observed in these galaxies in a variety of different wavelengths \citep{Weiss13, Spilker16, Reuter20}. \citet{Aravena16} used observations of CO(1-0) emission to determine the total molecular gas masses of both objects. Previous studies of these two sources used data with limited angular resolution and focused primarily on parametric lens mass modeling. Here we use data from a study designed to observe CO(3--2) in a selection of strongly gravitationally lensed galaxies. The original goal was to resolve down to scales of a few hundred parsecs in the source plane to compare any giant molecular clouds (GMCs) detected with local galaxy molecular gas scaling relations. We used these high-resolution observations to improve previous lens models, perform nonparametric modeling of the source plane emission, and, thus, the study the source-plane morphology of SPT\,0125-47 and SPT\,2134-50, specifically focusing on using nonparametric source plane lens modeling.

In Section \ref{sec:observation_details}, we describe the observations and data reduction steps. We present the results of the continuum, line analysis, and lens modeling in Section \ref{sec:results}. We provide a discussion of our results in Section \ref{sec:discussion}. Our conclusions are presented in Section \ref{sec:conclusions}. Throughout this paper, we adopt a flat $\Lambda$ cold dark matter ($\Lambda$CDM) cosmology with ${H_{0}}$ = 70\,km $\mathrm{s^{-1}\,Mpc^{-1}}$ and $\mathrm{\Omega_{m}}$ = 0.3.

\section{Observations and data reduction} \label{sec:observation_details}

SPT-S J012506-4723.7 (hereafter, SPT\,0125-47) and SPT-S J213403-5013.4 (hereafter SPT\,2134-50) were observed in band 3 as part of ALMA project ID 2016.1.01231.S (P.I. G. Drouart). The calibration and image processing steps were all performed in the Common Astronomy Software Application package \citep[CASA;][]{CASA}. The data for both sources were processed with the ALMA calibration pipeline in CASA 4.7.2, which includes the calibration of the phase, bandpass, flux, and gain. For SPT\,0125-47, J2357-5311 was used as the bandpass and flux calibrator, J0124-5113 was used as the phase calibrator, and J0133-4430 was used as the check source. For SPT\,2134-50, J2056-4714 was used as the bandpass and flux calibrator, J2124-4948 was used as the phase calibrator, and J2135-5006 was used as the check source. The pipeline-reduced data and diagnostic plots were visually inspected to ensure data quality, and additional flagging was added where necessary to ensure the calibration was satisfactory. The pipeline-reduced measurement sets were used to create images using the CASA task {\sc TCLEAN} with a Hogbom deconvolution using private scripts in CASA version 5.6.2-3, where cleaning was performed in the region of expected emission. These images were used to identify line-free channels for continuum subtraction. The continuum was subtracted using the CASA task {\sc UVCONTSUB} with a polynomial fit of order 1 for both SPT\,0125-47 and SPT\,2134-50. A natural weighting scheme was used when imaging the dust continuum emission. Emission line cubes were cleaned down to $1\sigma$ levels using {\sc TCLEAN} and imaged using a natural weighting scheme with a spectral resolution of $\sim$70\,km\,s$^{-1}$, where $1\sigma = 0.5$\,mJy for SPT\,0125-47 and $1\sigma = 0.36$\,mJy for SPT\,2134-50. The residual images were checked for emission to ensure that the cleaning was adequate. The maximum recoverable scale was $\approx 15.2''$ for SPT\,0125-47 and $\approx 16.3''$ for SPT\,2134-50. We conservatively report the uncertainty in the absolute flux calibration to be $\sim 10\%$ (note that the fiducial value in band 3 is 5\%)\footnote{https://almascience.eso.org/documents-and-tools/cycle10/alma- technical-handbook}. Observational details for both sources are provide in Table \ref{tab:obs_details}. 

\section{Results} \label{sec:results}
\subsection{Continuum emission} \label{subsec:continuum_results}

\begin{figure}
    \centering
    \includegraphics[width = 0.75\linewidth]{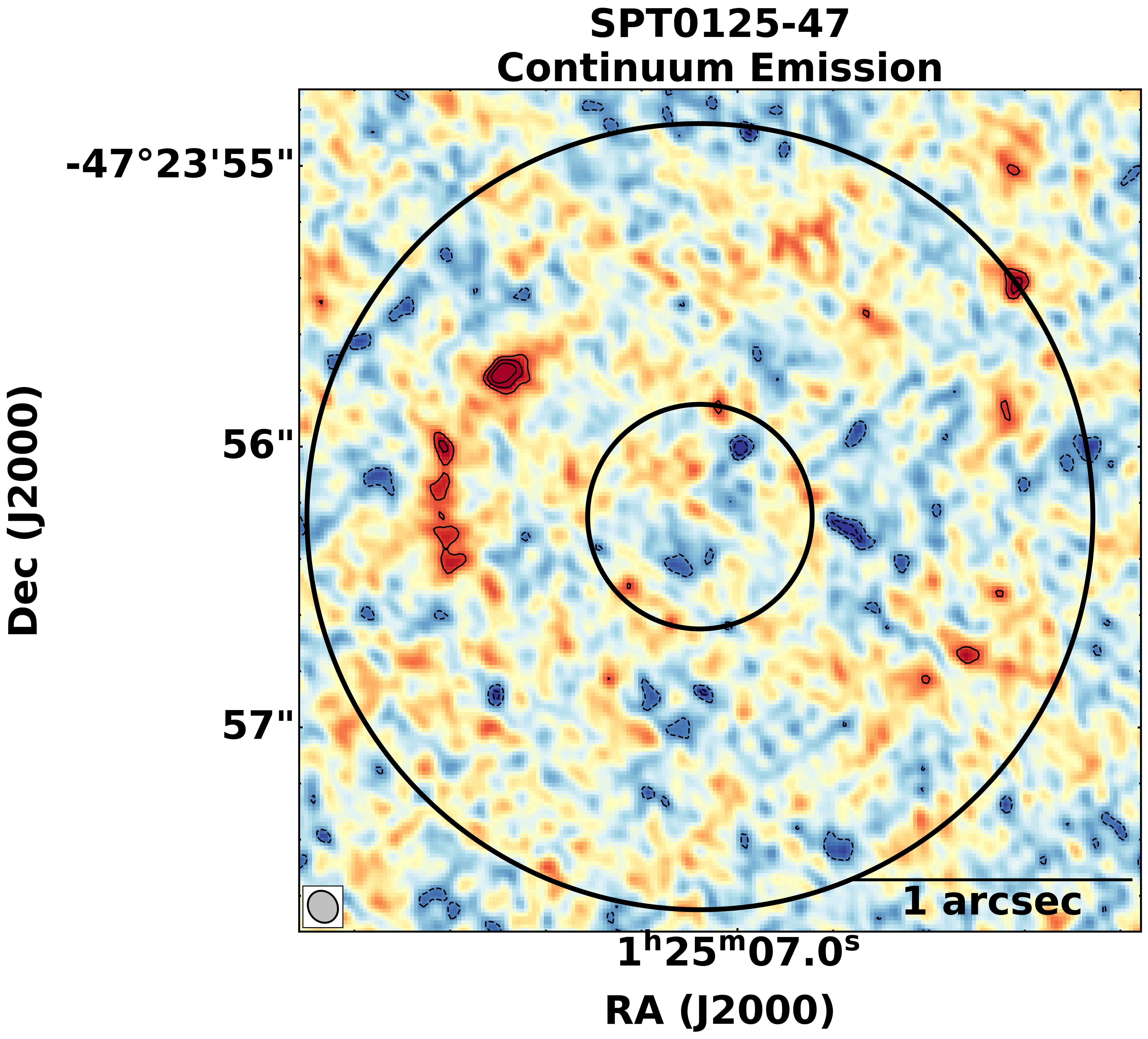}
    \includegraphics[width = 0.75\linewidth]{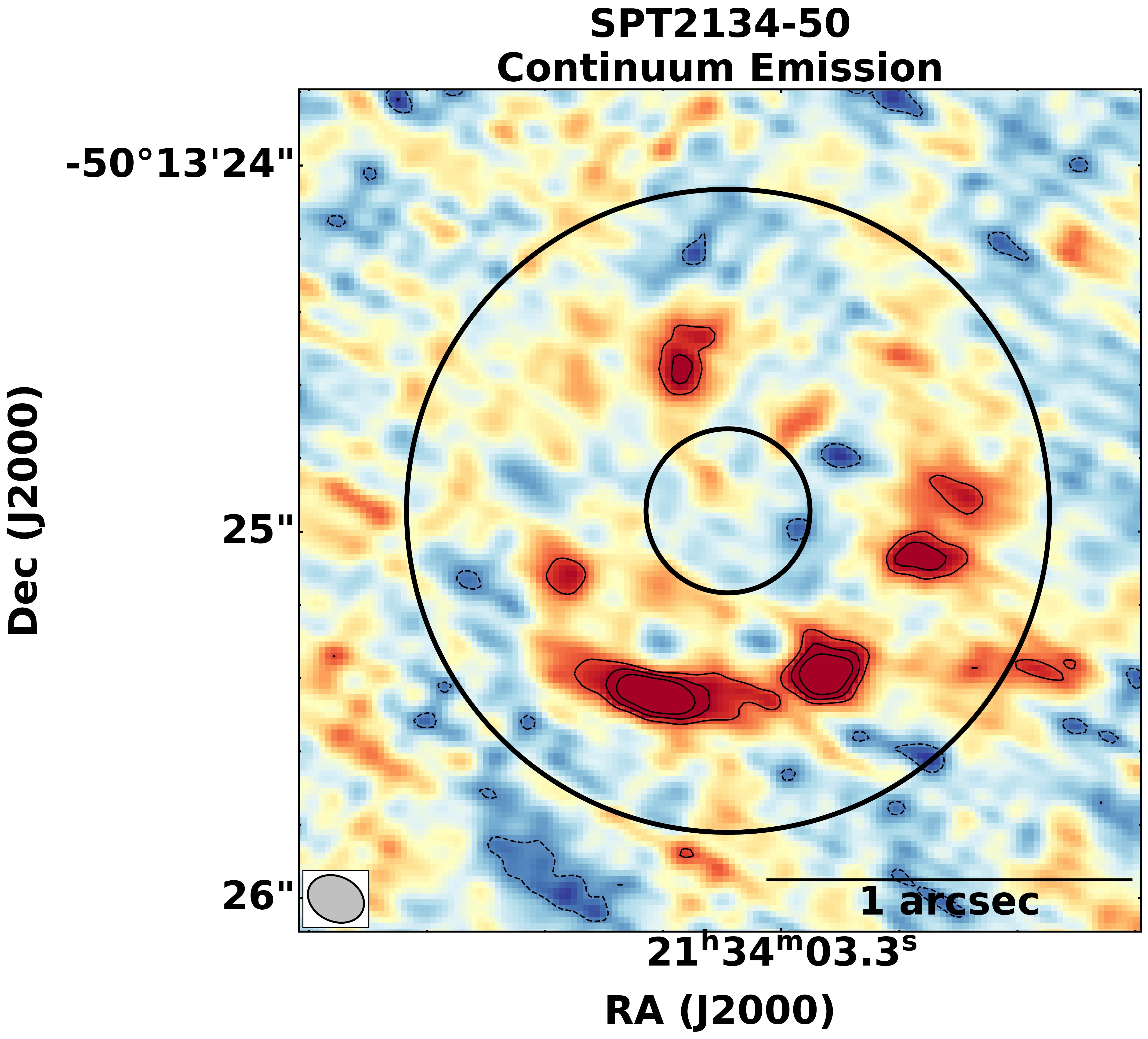}
    \caption{Continuum image of SPT\,0125-47 (top) and SPT\,2134-50 (bottom). The contours are shown at $-3, -2, 3, 4,  5\sigma$ levels. The synthesized beam is shown in the bottom left of each image. The black annulus shows the regions used to extract the dust continuum emission.}
    \label{fig:continuum}
\end{figure}

We extracted the dust continuum emission for both sources from emission regions above $3\sigma$ within a chosen circular annular aperture shown in Fig. \ref{fig:continuum}. We note that for both sources the continuum flux density is lower than the CO(3--2) emission. This is expected as the observations used in this work probe the longer wavelength regime and thereby the colder dust. We detect dust continuum emission toward SPT\,0125-47 at $1.9 \pm 1.4$\,mJy (not corrected for lensing) and toward SPT\,2134-50 at $1.1 \pm 0.7$\,mJy (not corrected for lensing). Both values are in good agreement with the value reported in \citet{Weiss13}\footnote{Given that these continuum measurements are in good agreement with previous measurements at the same wavelength, we do not pursue spectral energy distribution (SED) fitting.}. We show the continuum images for both SPT\,0125-47 and SPT\,2134-50 in Fig. \ref{fig:continuum}.

\subsection{Line emission} \label{subsec:lineemission_results}

\begin{figure*}
    \centering
    \includegraphics[width = 1.0\linewidth]{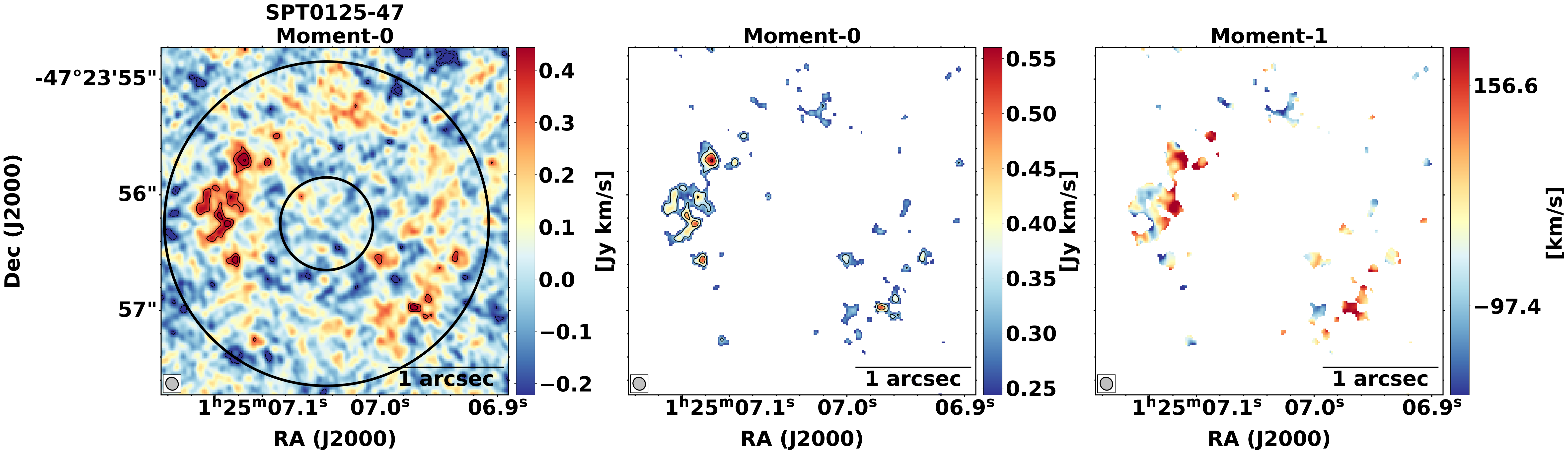}
    \includegraphics[width = 1.0\linewidth]{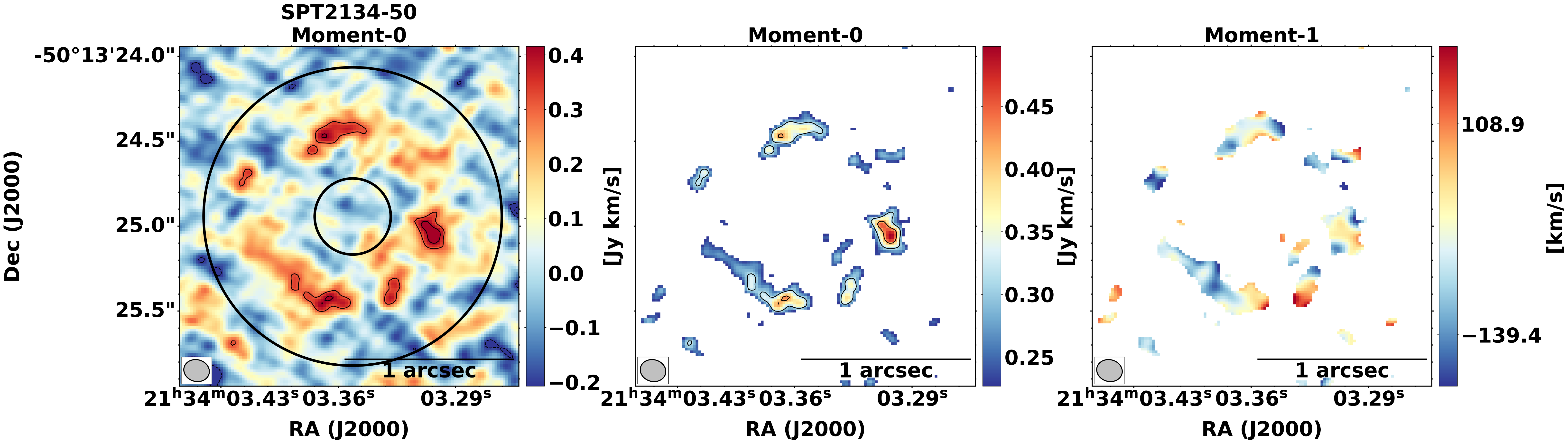}
    \caption{Moment-0 and moment-1 maps of the CO(3--2) emission for SPT\,0125-47 (top) and SPT\,2134-50 (bottom). The first column shows the unmasked moment-0 map, the second column shows the moment-0 map masked to show only values above $3\sigma$, and the third column shows the moment-1 map. The contours in the first two columns are shown at $-3, -2, 3, 4, 5, 6, 7, 8, 9, 10\sigma$ levels. The synthesized beam is shown in the bottom left of each image. The black annulus in the unmasked moment-0 map for each source shows where the spectrum was extracted from.}
    \label{fig:mom0_1}
\end{figure*}

\begin{figure*}
    \centering
    \includegraphics[width = 0.4\linewidth]{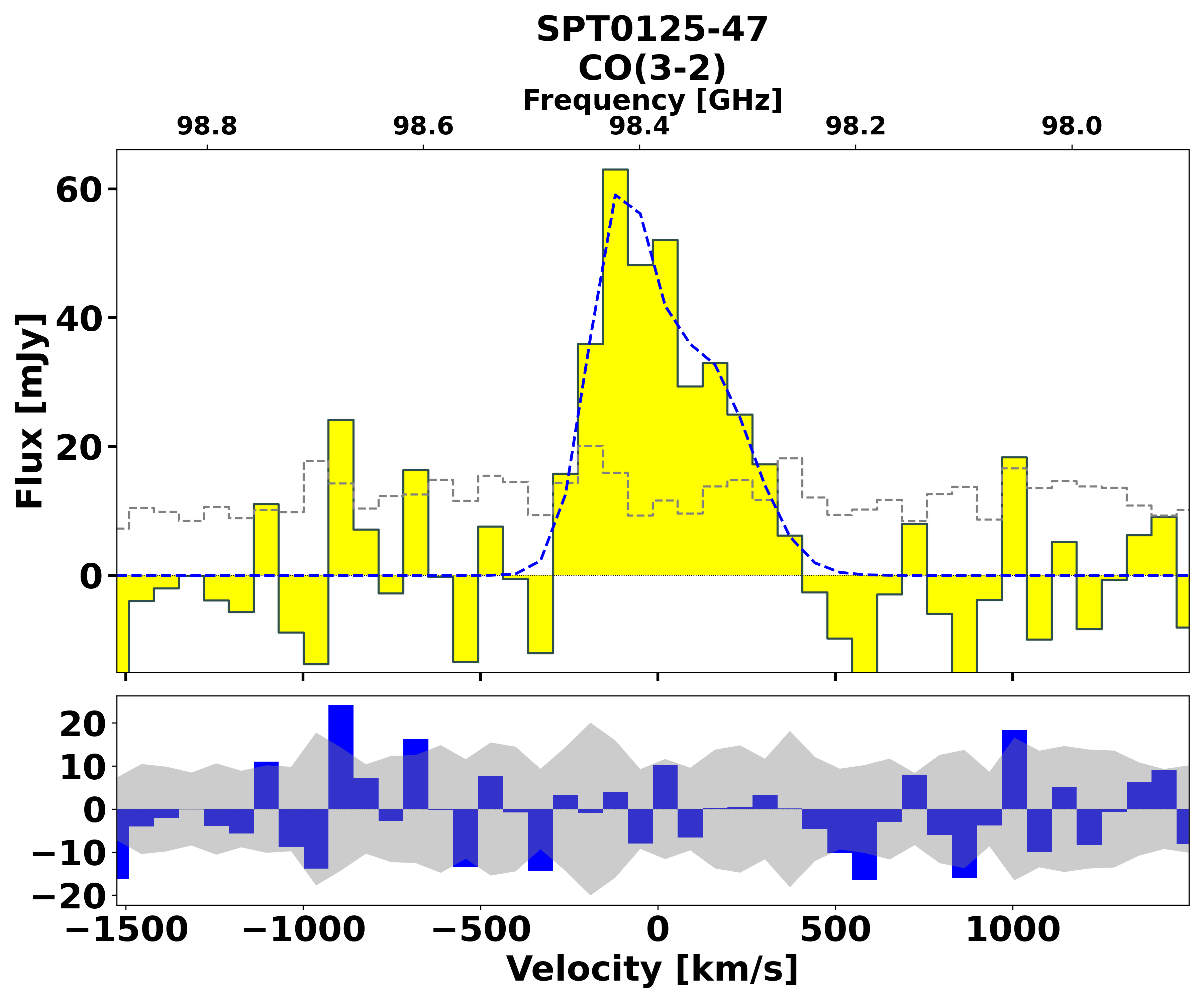}
    \includegraphics[width = 0.4\linewidth]{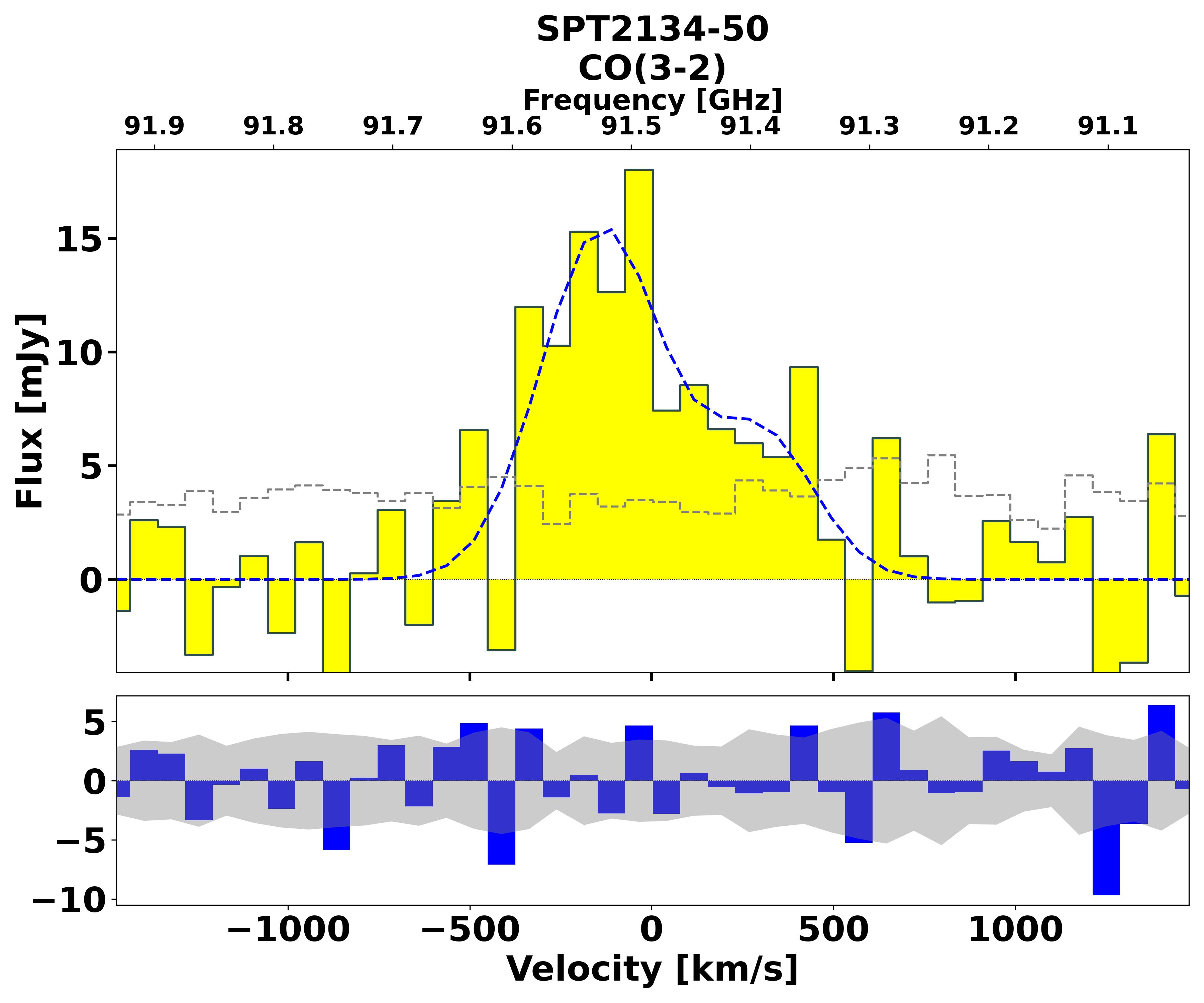}
    \caption{Spectra of the CO(3--2) toward SPT\,0125-47 (left) and SPT\,2134-50 (right). In both cases, the spectrum is shown in the top panel, and the residuals from the Gaussian fit are shown in the bottom panel. The dashed blue line shows the two Gaussian fit to the spectra. The dashed gray line in the top panel and the shaded gray region in the lower panel indicate the per-channel rms. Additionally, the top axis in the top panel of each spectrum displays the corresponding frequency.}
    \label{fig:spectra}
\end{figure*}

We detect CO(3--2) (rest frequency 345.7959899\,GHz) emission toward both SPT\,0125-47 and SPT\,2134-50. We performed a regional extraction using a circular annular region (the same as that used for the continuum) of both spectra without including a $\sigma$ cut to the data. The region of extraction is shown in Fig. \ref{fig:mom0_1}. This methodology preserves fainter flux at emission edges, flux that a $5\sigma$ cut would exclude, making it particularly useful for analyzing "clumpy" structures. This approach was advantageous given the high angular resolution of the observations for both sources. To calculate the root mean square (rms) in each channel of the spectrum, we employed a straightforward sampling method. This involved sampling multiple emission-free regions of the cube in each channel, with the sampled regions matching the size of those used for spectrum extraction. The spectra of both sources are shown in Fig. \ref{fig:spectra}. Due to the skewed profile of both emission line profiles, we fit the CO(3--2) emission with two Gaussian profiles (a single Gaussian does not provide a good fit to the data). We created moment-0 and moment-1 maps across the emission line, shown in Fig. \ref{fig:mom0_1}. Line properties are provided in Table \ref{tab:Line_properties}. Integrated flux densities and line luminosities were calculated using the following equations from \citet{Solomon97}:\ 

\begin{equation}
    L_{\rm line} = (1.04 \times 10^{-3})\,I_{\rm obs}\, \nu_{\rm rest}\,D^{2}_{L}\,(1+\textit{z})^{-1} [L_{\odot}],
\end{equation}

where $L_{\rm line}$ [L$_{\odot}$] is the luminosity of the emission line, $I_{\rm obs}$ is the velocity integrated flux density ($I_{\mathrm{obs}} = S_{\mathrm{line}} \Delta V$ [Jy\,km\,s$^{-1}$]), $S_{\mathrm{line}}$ [mJy] is the observed flux density, and $\Delta V$ [km\,s$^{-1}$] is the full width at half maximum (FWHM) of the emission line, $\nu_{\rm rest}$ [GHz] is the rest-frame frequency of the line, $D_{\rm L}$ [Mpc] is the luminosity distance and $z$ is the redshift. This is expressed as 

\begin{equation}
    L^{'}_{\rm line} = (3.25 \times 10^7)\,I_{\rm obs}\,D^{2}_{L}\,(1+\textit{z})^{-3}\, \nu_{\rm obs}^{-2} [\rm K \, km\, s^{-1} pc^{-2}], 
\end{equation}
where $L^{'}_{\rm line}$ is the luminosity of the emission line, $I_{\rm obs}$ is the velocity integrated flux density ($I_{\mathrm{obs}} = S_{\mathrm{line}} \Delta V$ [Jy\,km\,s$^{-1}$]), $D_{L}$ [Mpc] is the luminosity distance of the source, $\nu_{\rm rest}$ [GHz] is the rest-frame frequency of the line, and $z$ is the redshift.

\begin{table*}[h]
    \centering
    \caption{Line properties from our observations.} 
    \begin{tabular}{l c c c c c c c c} \hline \hline
        Source & z & $S_{v, \rm peak}^{a,g}$ & FWHM$_{\rm peak}^{b}$ & $I_{\mathrm{mol}}^{c,g}$ & $L_{\rm mol}^{d,g}$ & $L_{\rm mol}^{',e,g}$ & $\boldsymbol{\mu_{CO(3-2)}} ^{f}$ \\  
         & & [mJy] & [km\,$\rm s^{-1}$] & [Jy km $\rm s^{-1}$] & [$10^{8}\,L_{\odot}$] & [$10^{11}$\,K\,km\,s$^{-1}$] & \\ \hline
        
        SPT\,0125-47 & 2.5148 & $5.0 \pm 2.5$ & $206 \pm 73$ & $2.1 \pm 1.0$ & $0.96 \pm 0.50$ & $0.73 \pm 0.35$ & $10.7 \pm 0.002$ \\
        & & $3.1 \pm 1.0$ & $310 \pm 212$ & & & \\

        SPT\,2134-50 & 2.7799 & $2.0 \pm 0.3$ & $394 \pm 128$ & $1.2 \pm 0.4$ &  $0.62 \pm 0.22$ & $0.47 \pm 0.17$ & $7.6 \pm 0.002$ \\
        & &  $0.9 \pm 0.4$ & $348 \pm 271$  & & & \\ \hline
        
       \hline
    \end{tabular}
        \tablefoot{
        \tablefoottext{a}{Specific intensity of the two peaks comprising the double Gaussian fit.}
        \tablefoottext{b}{FWHM of the two components comprising the double Gaussian fit.}
        \tablefoottext{c}{Intensity of the entire line.}
        \tablefoottext{d}{Luminosity of the entire line expressed in $\rm L_{\odot}$.}
        \tablefoottext{e}{Luminosity of the entire line expressed in $\rm K\,km\,s^{-1}\,pc^{-2}$.}
        \tablefoottext{f}{Magnification factor for the entire line.}
        \tablefoottext{g}{Corrected for lensing magnification.}
        }
    \label{tab:Line_properties}
\end{table*}

\subsection{Lens modeling} \label{subsec:lens_modeling}

\begin{figure*}[h]
    \centering
    \includegraphics[width = 0.19\linewidth]{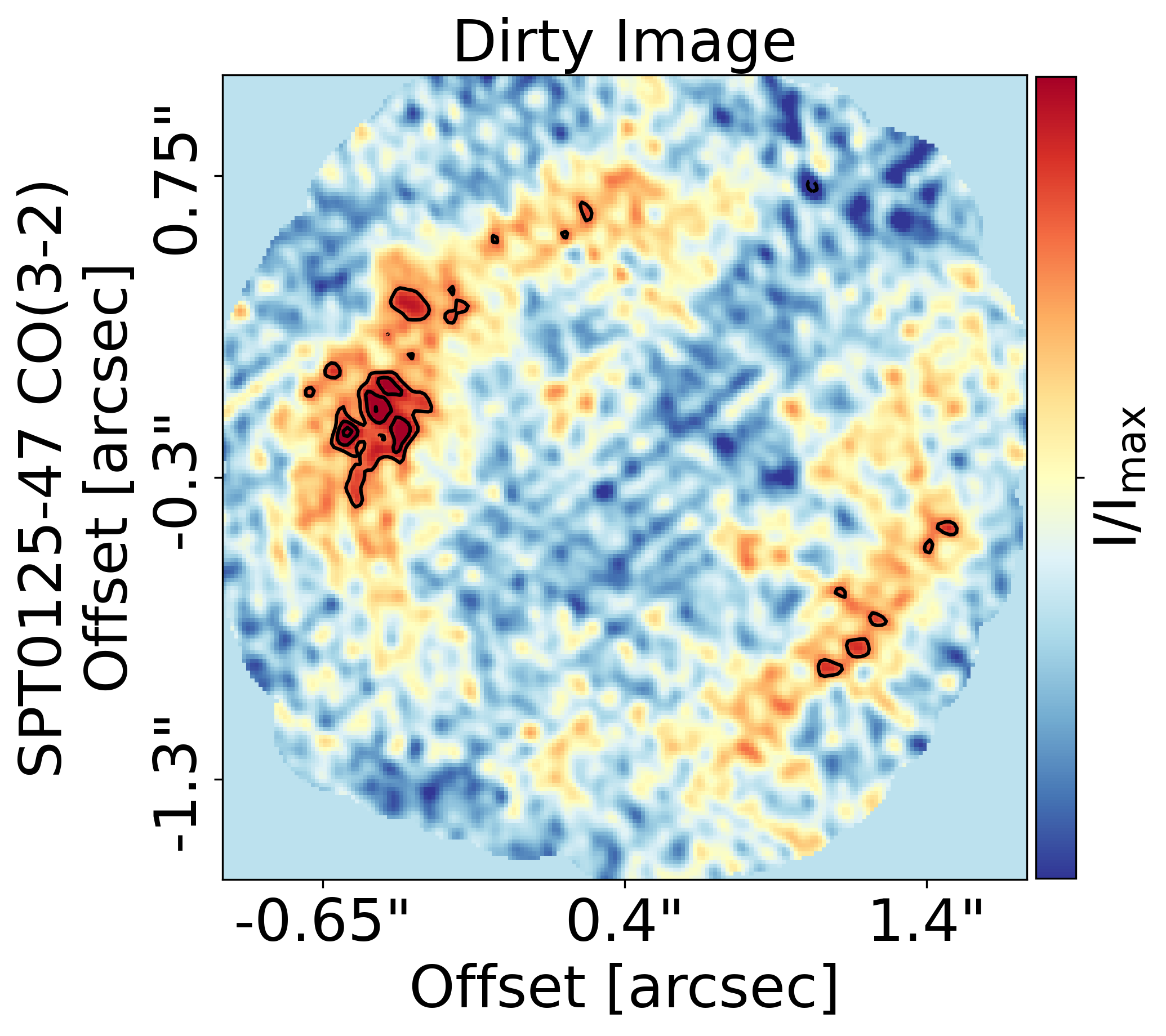}
    \includegraphics[width = 0.18\linewidth]{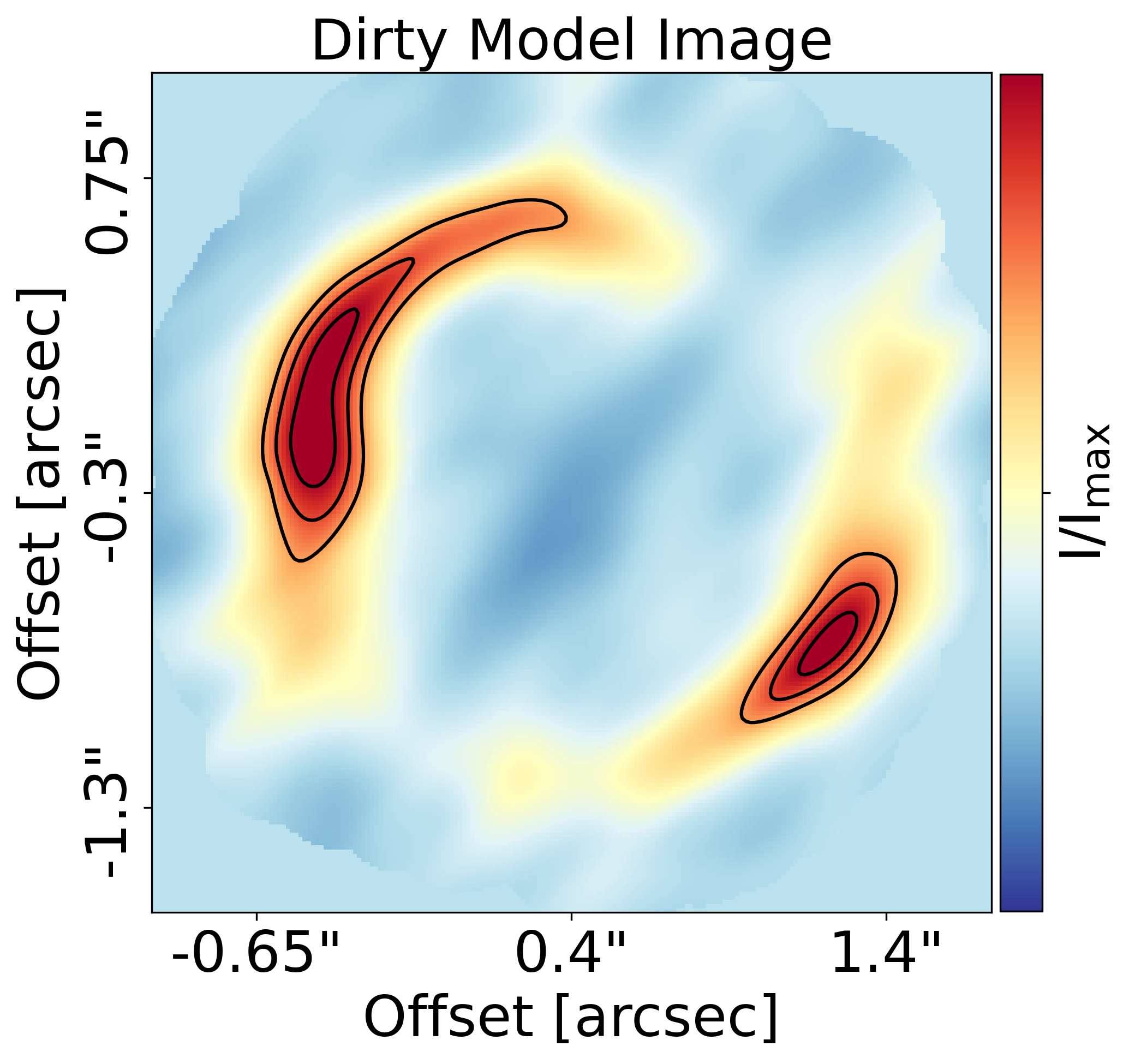}
    \includegraphics[width = 0.18\linewidth]{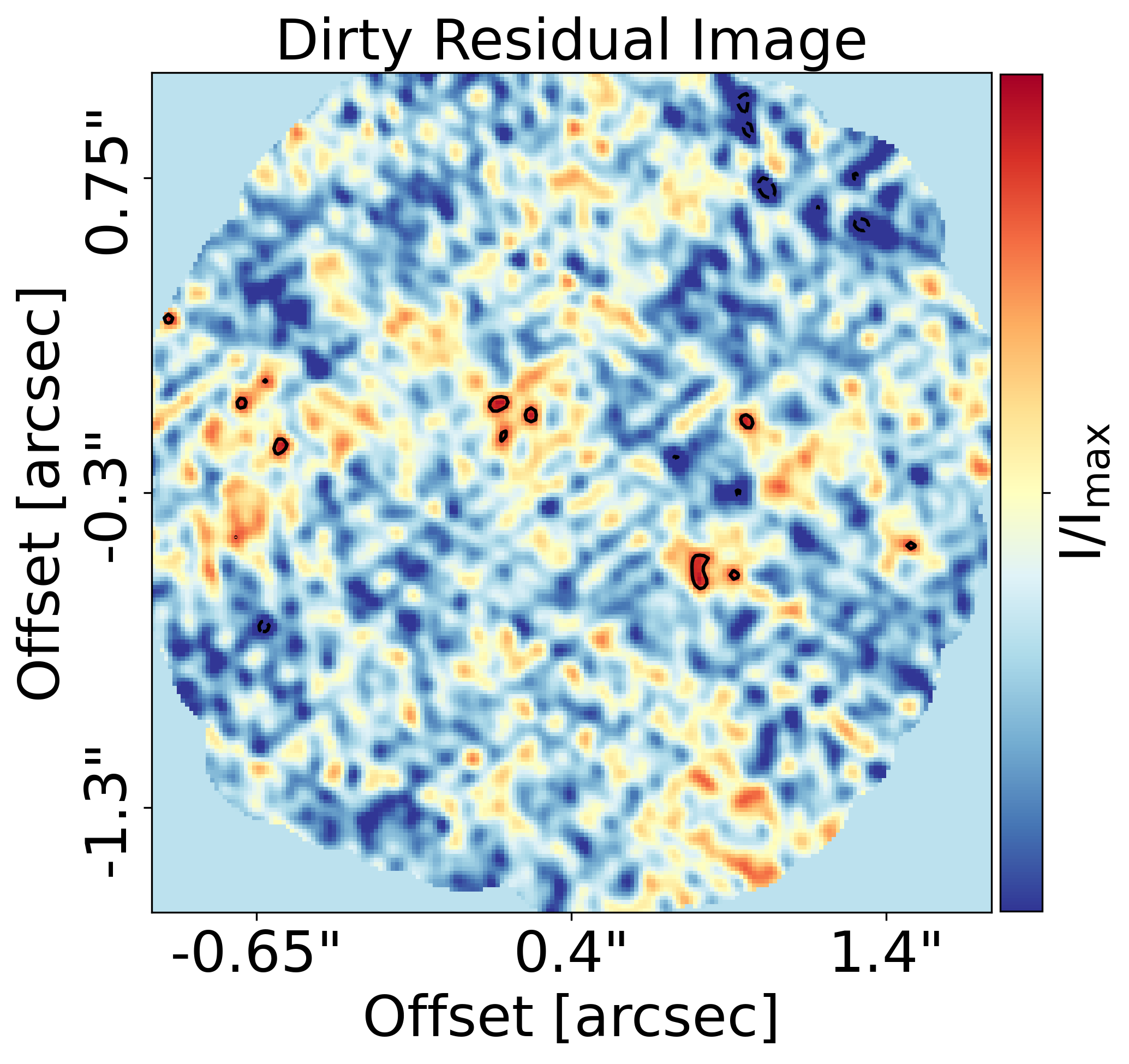}
    \includegraphics[width = 0.18\linewidth]{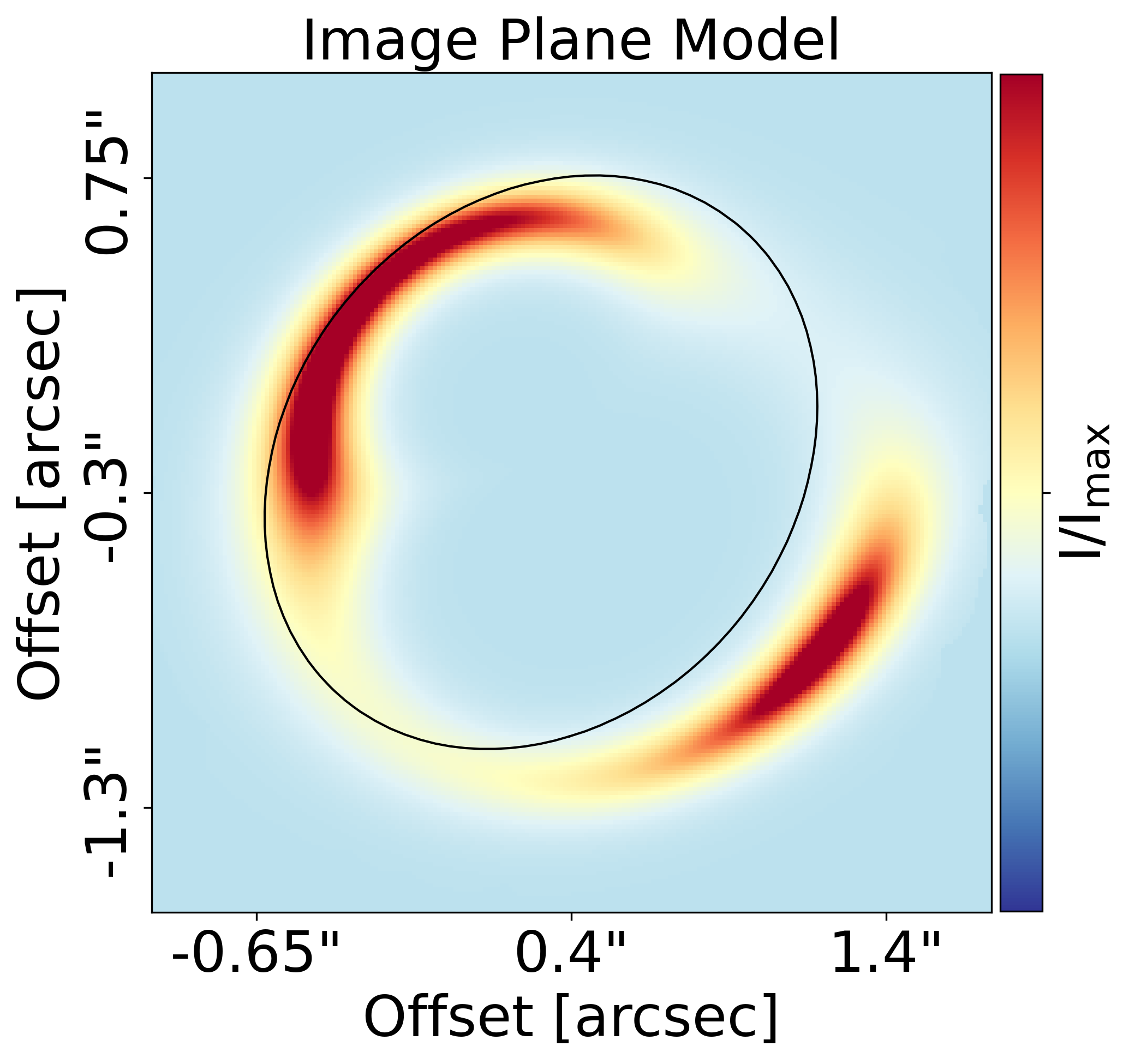}
    \includegraphics[width = 0.18\linewidth]{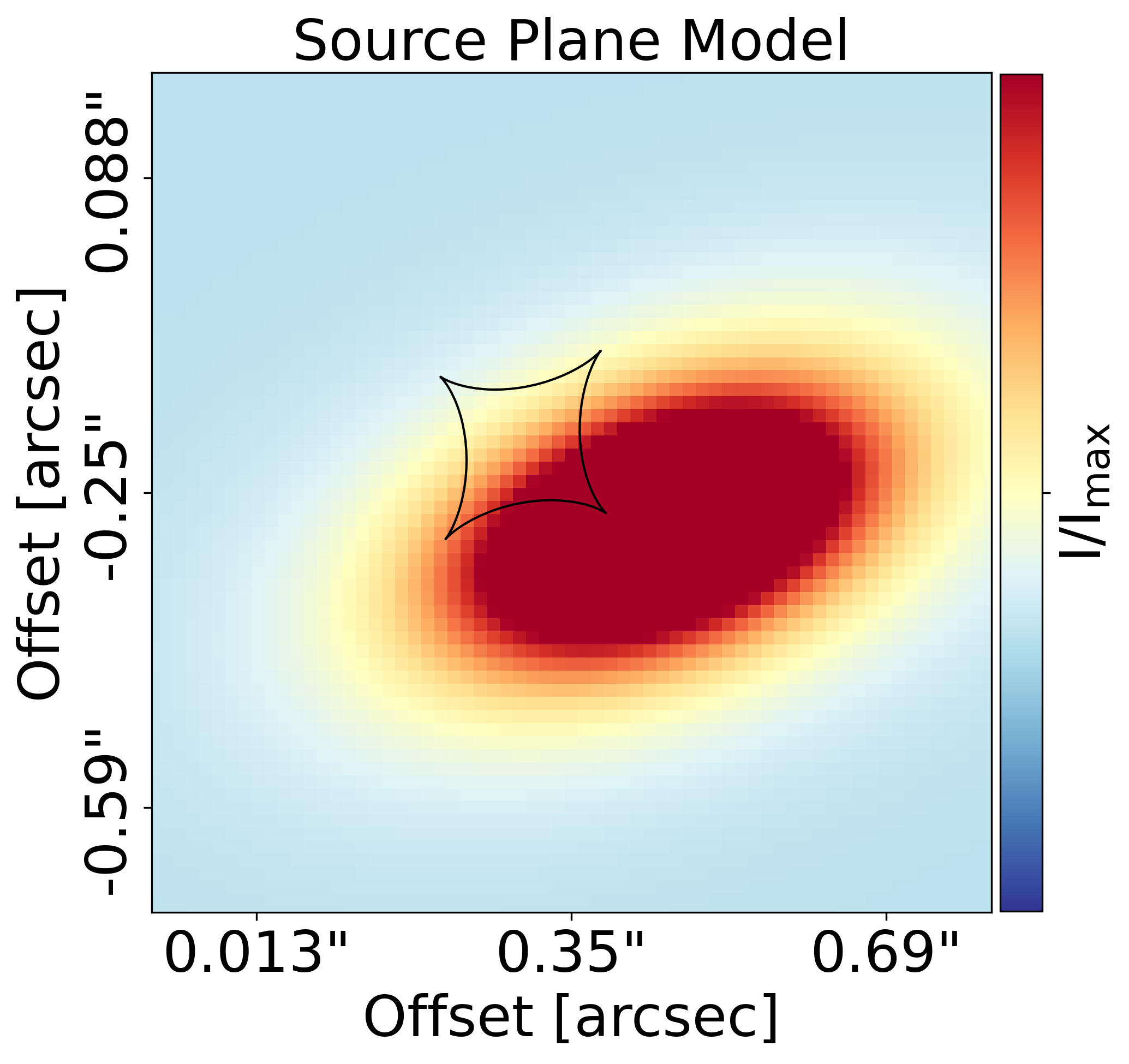}

    \includegraphics[width = 0.19\linewidth]{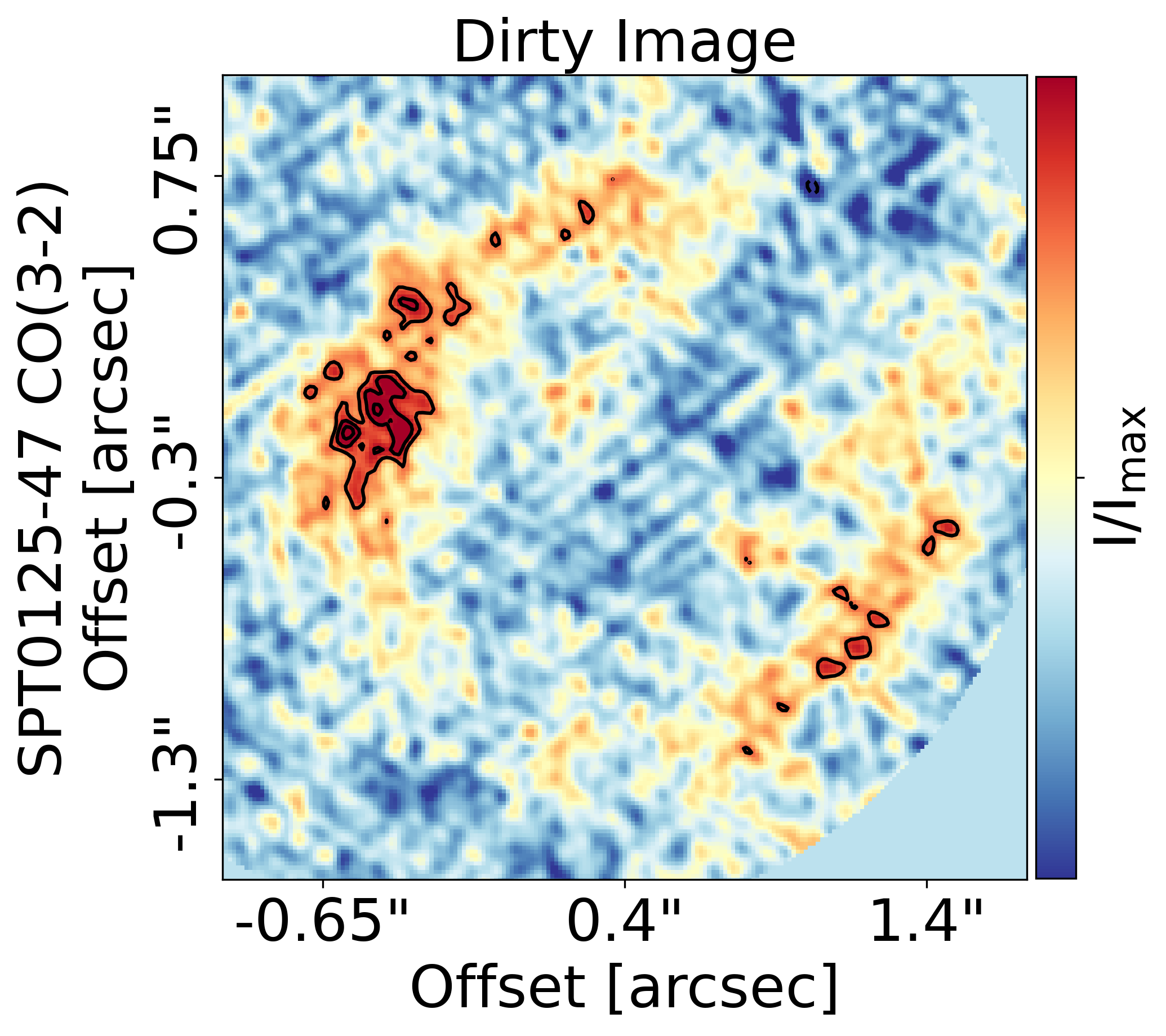}
    \includegraphics[width = 0.18\linewidth]{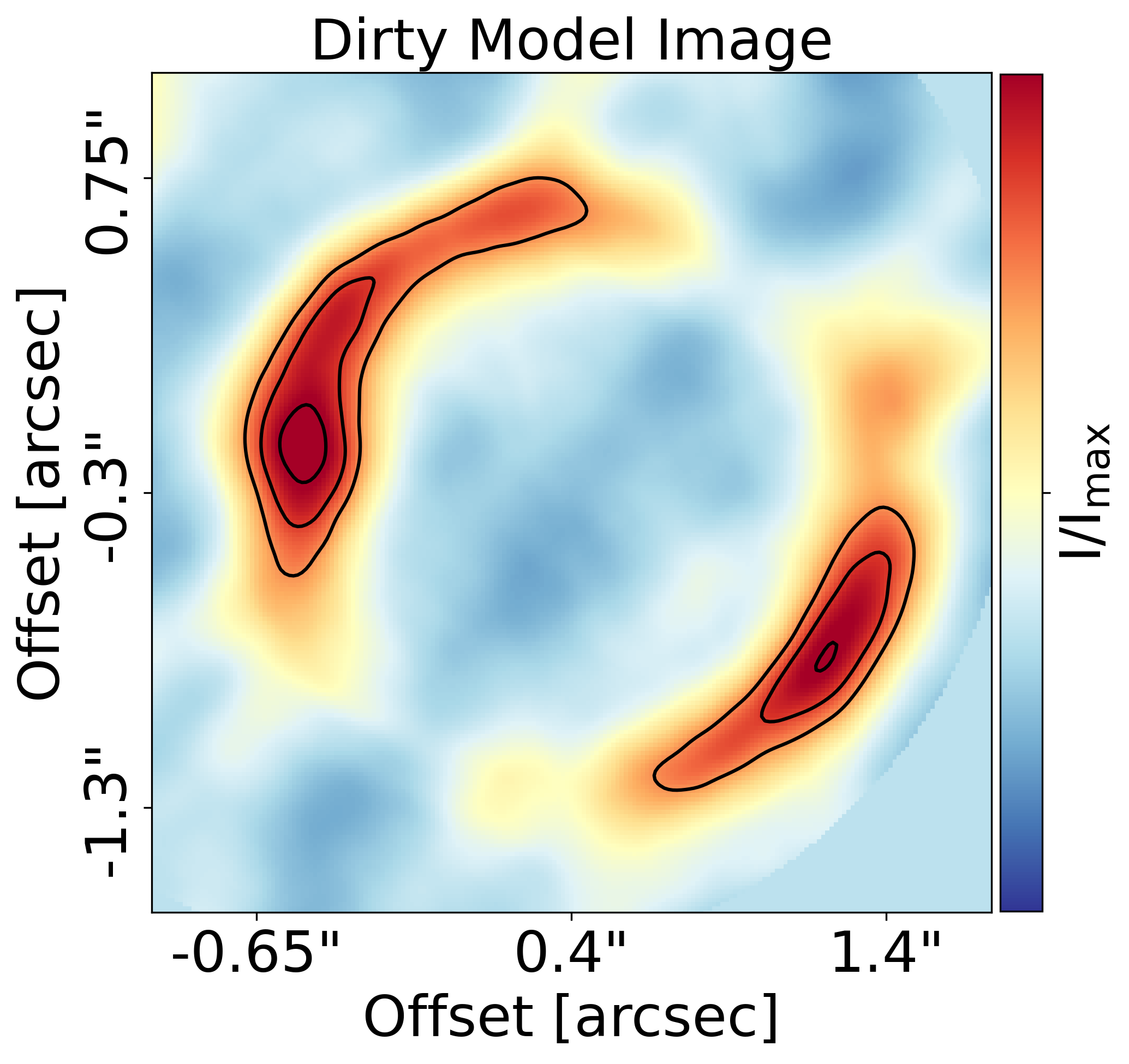}
    \includegraphics[width = 0.18\linewidth]{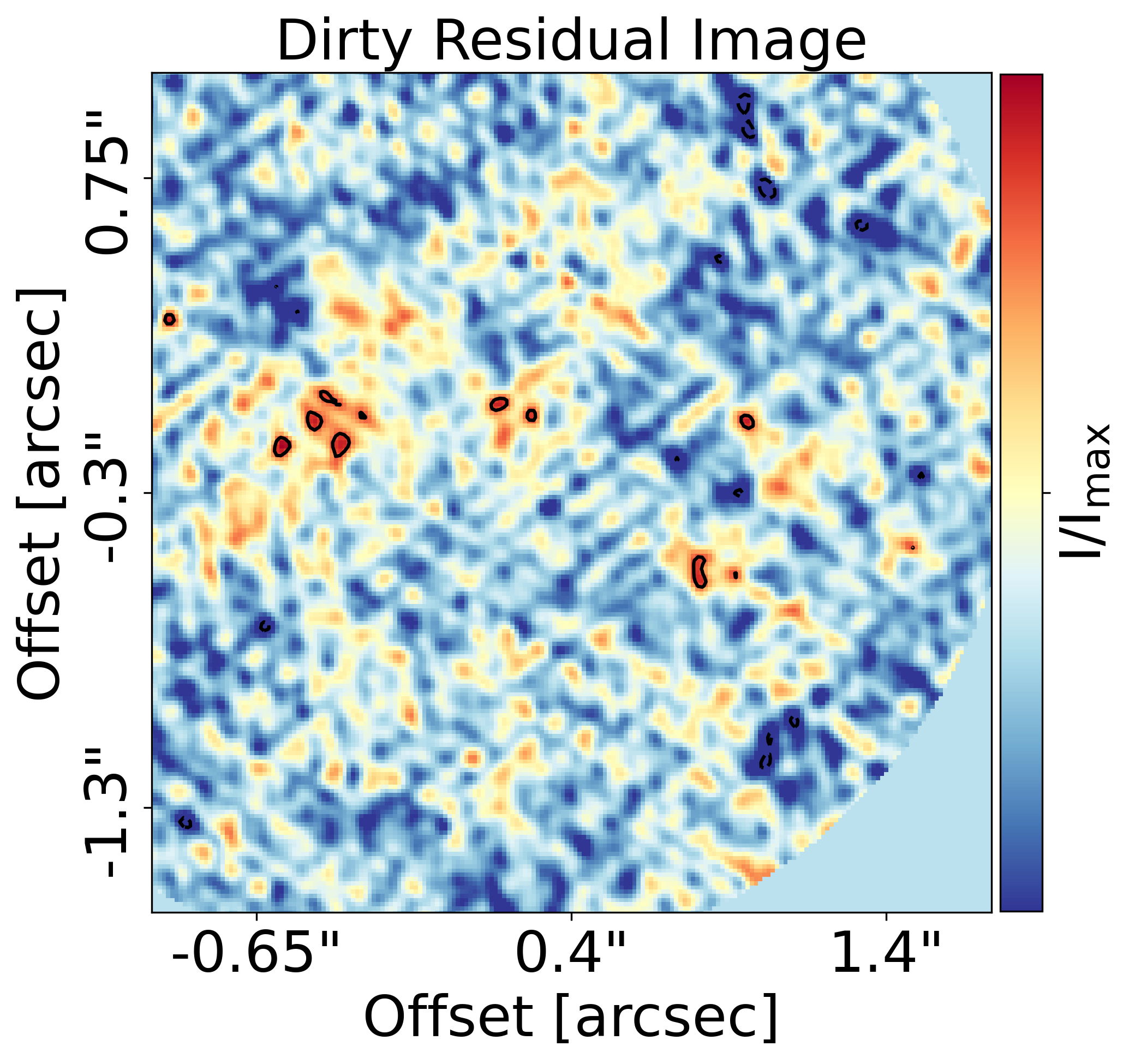}
    \includegraphics[width = 0.18\linewidth]{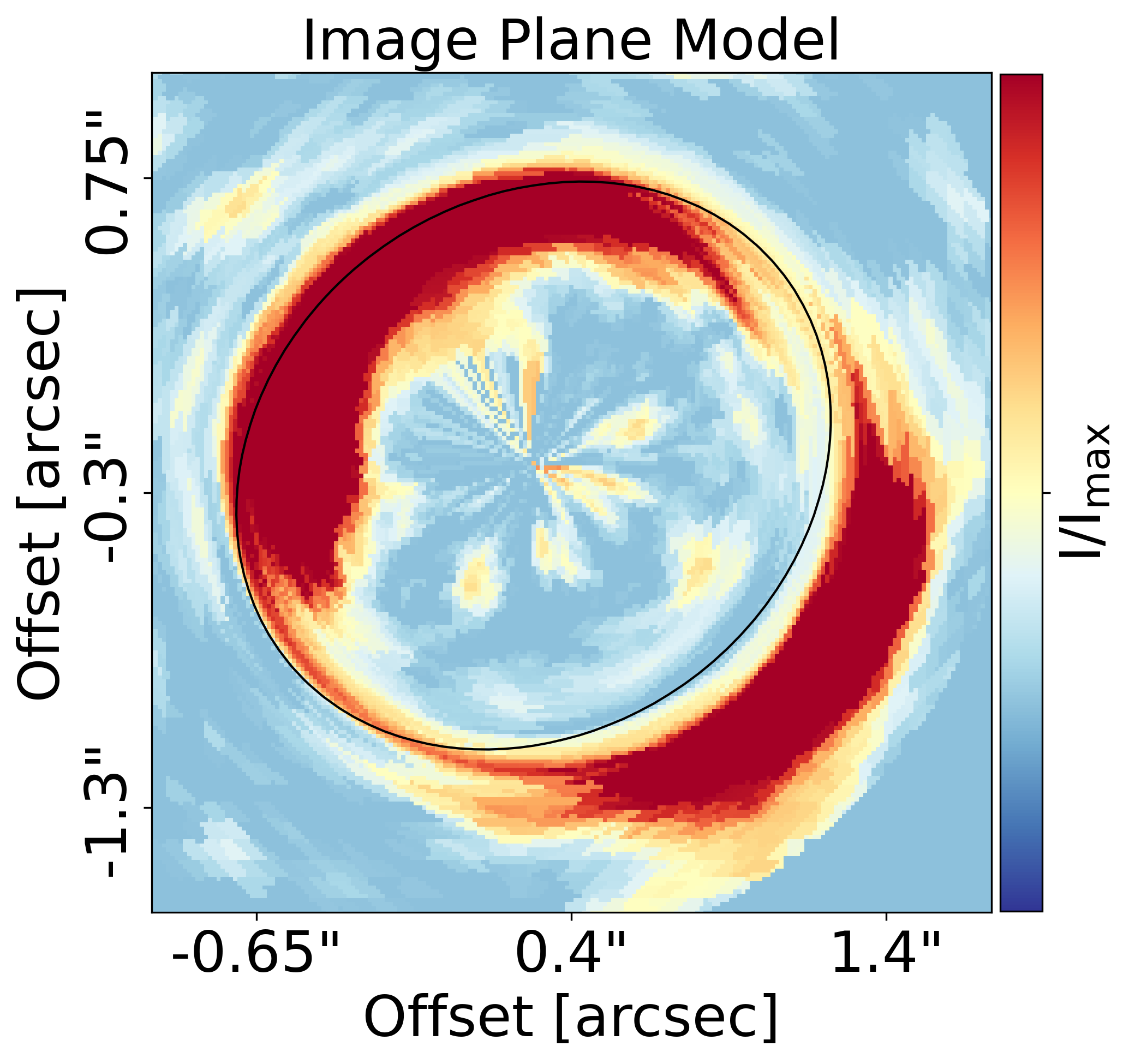}
    \includegraphics[width = 0.18\linewidth]{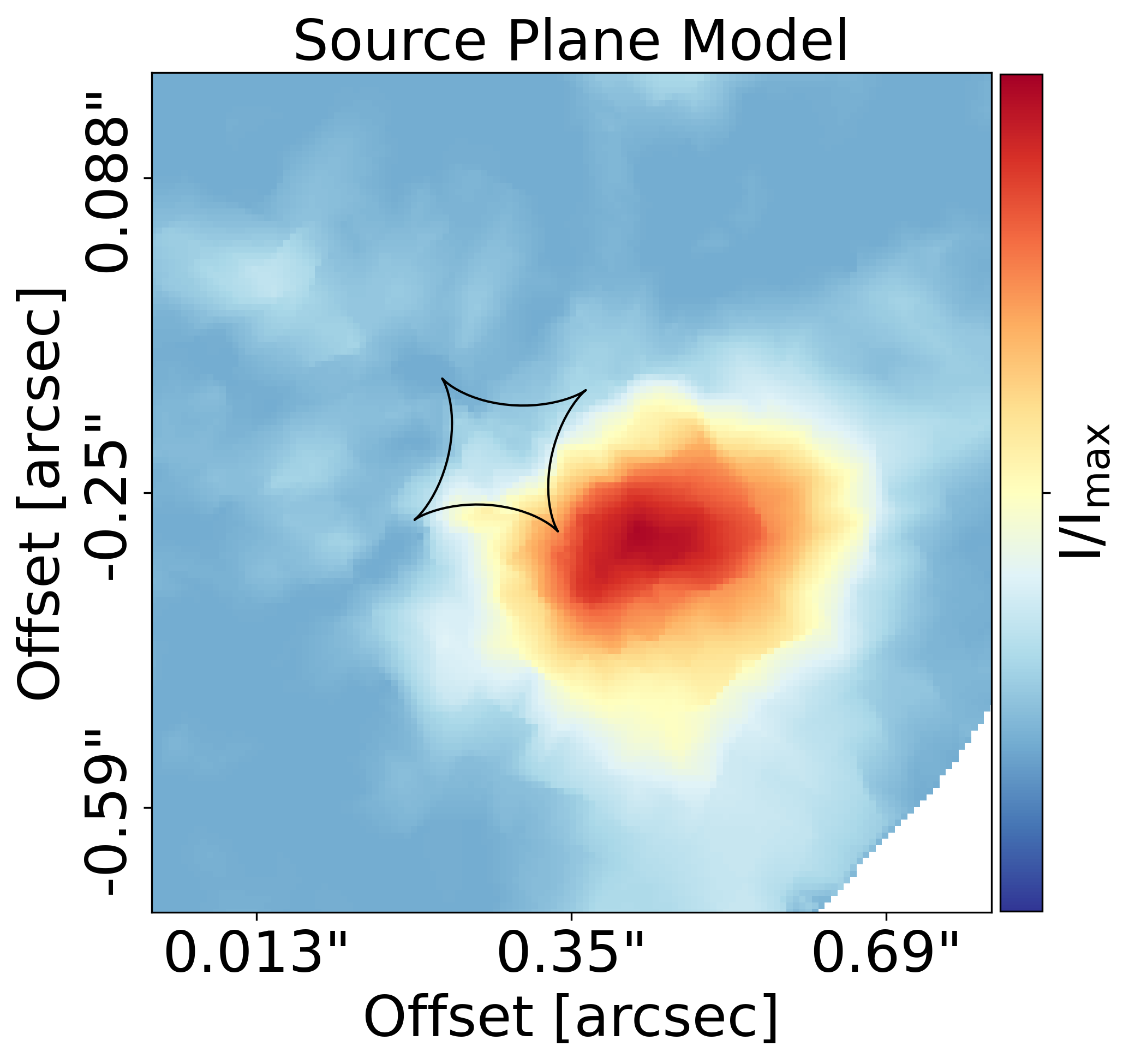}

    \includegraphics[width = 0.19\linewidth]{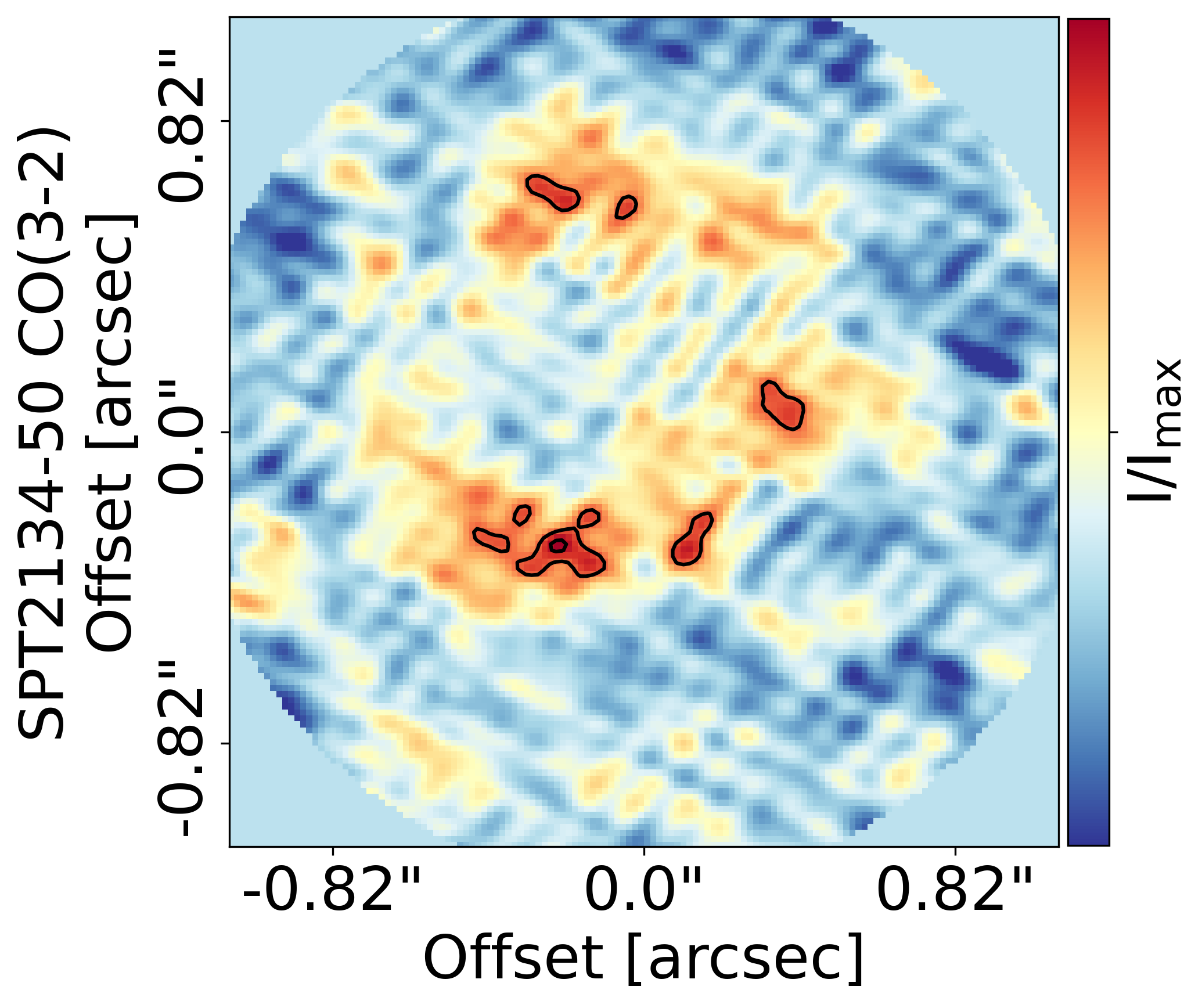}
    \includegraphics[width = 0.18\linewidth]{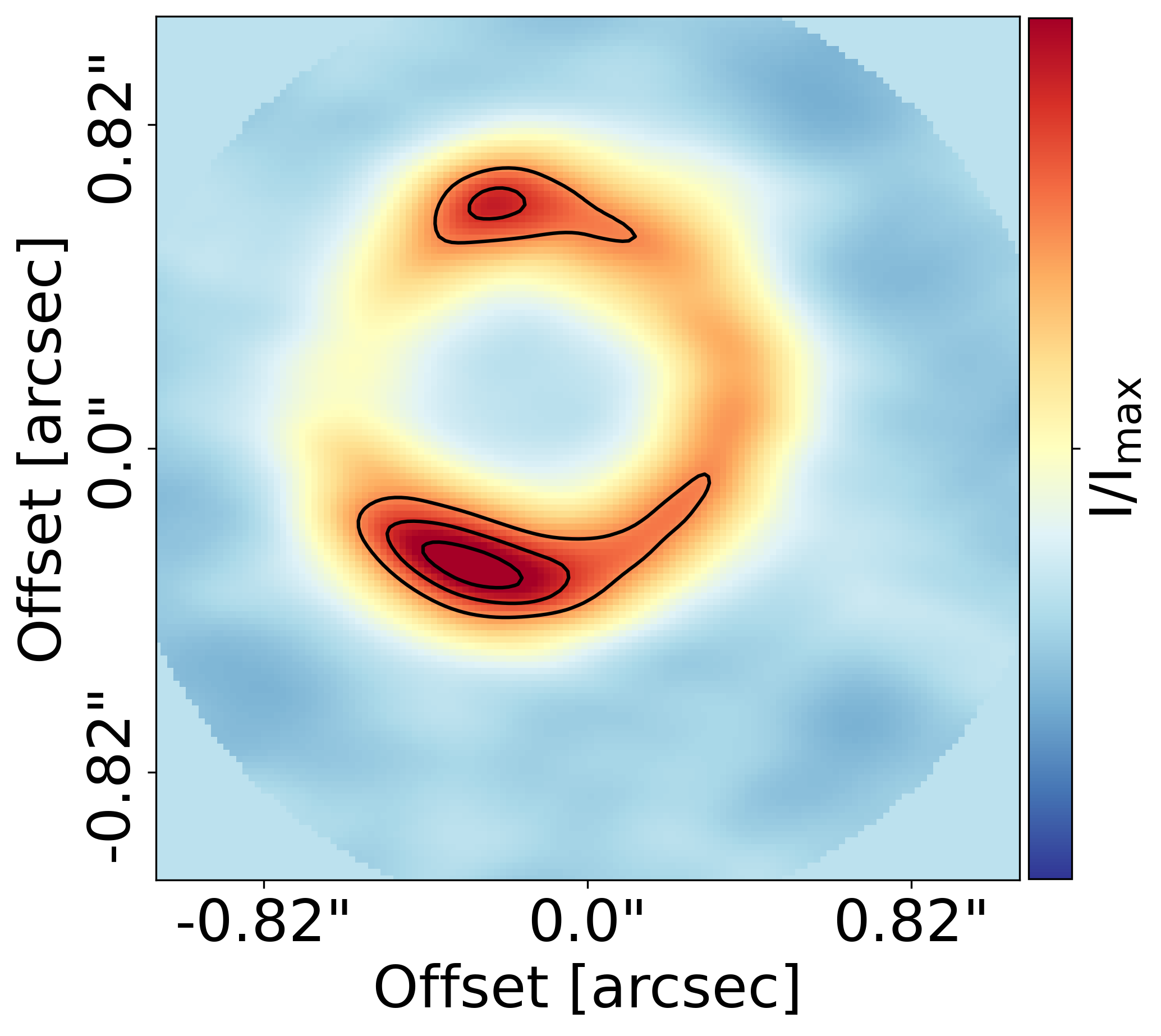}
    \includegraphics[width = 0.18\linewidth]{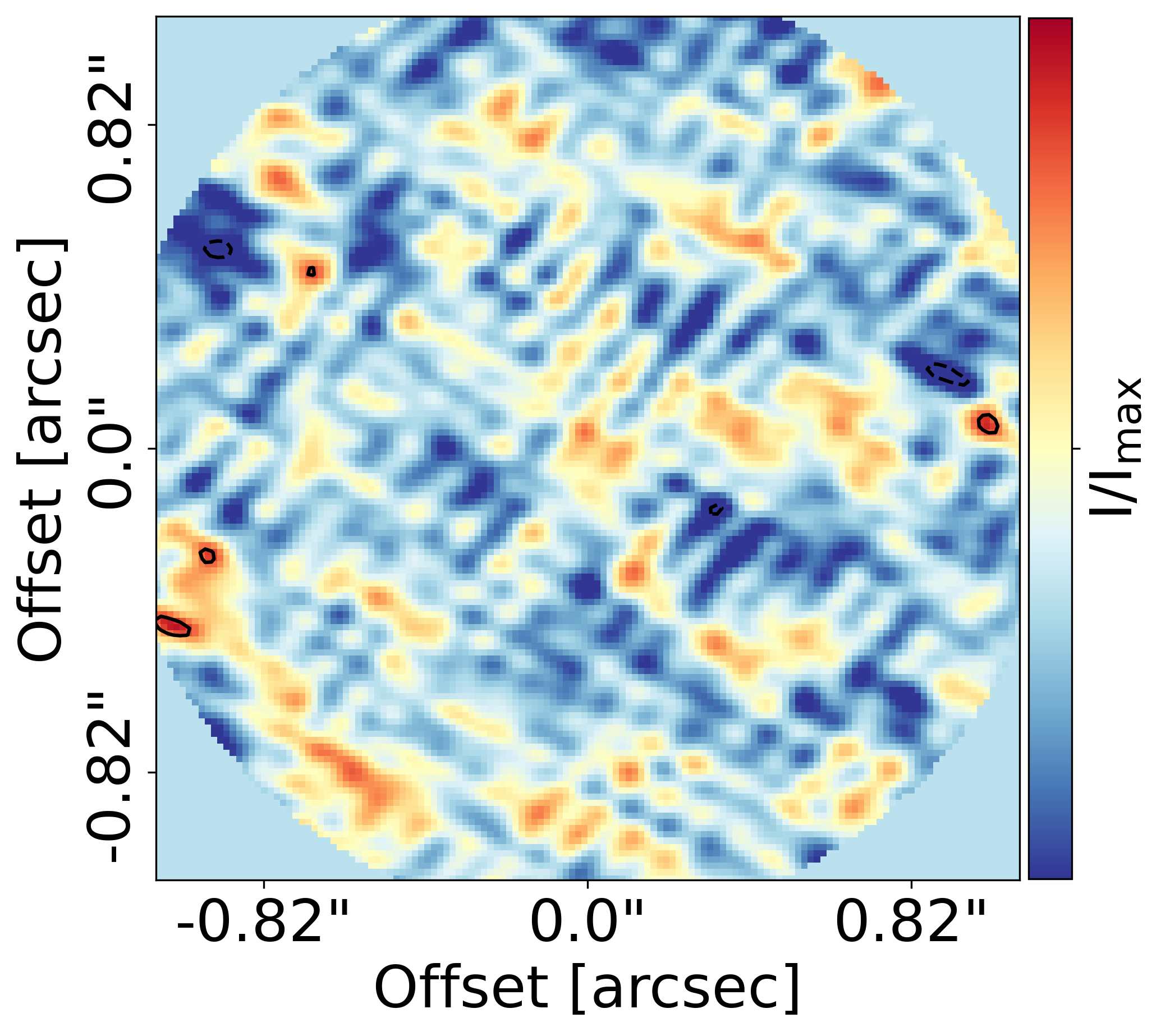}
    \includegraphics[width = 0.18\linewidth]{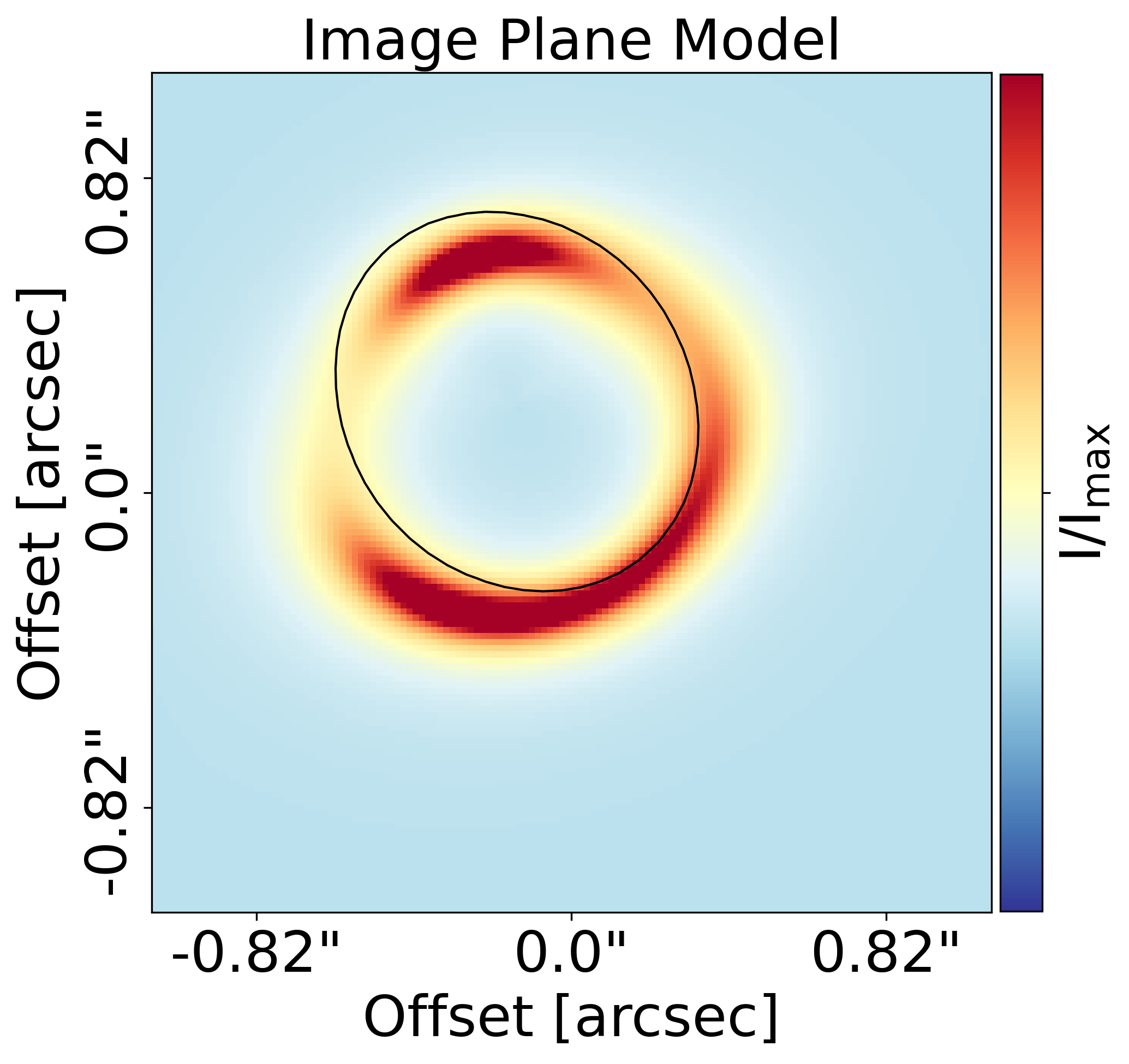}
    \includegraphics[width = 0.18\linewidth]{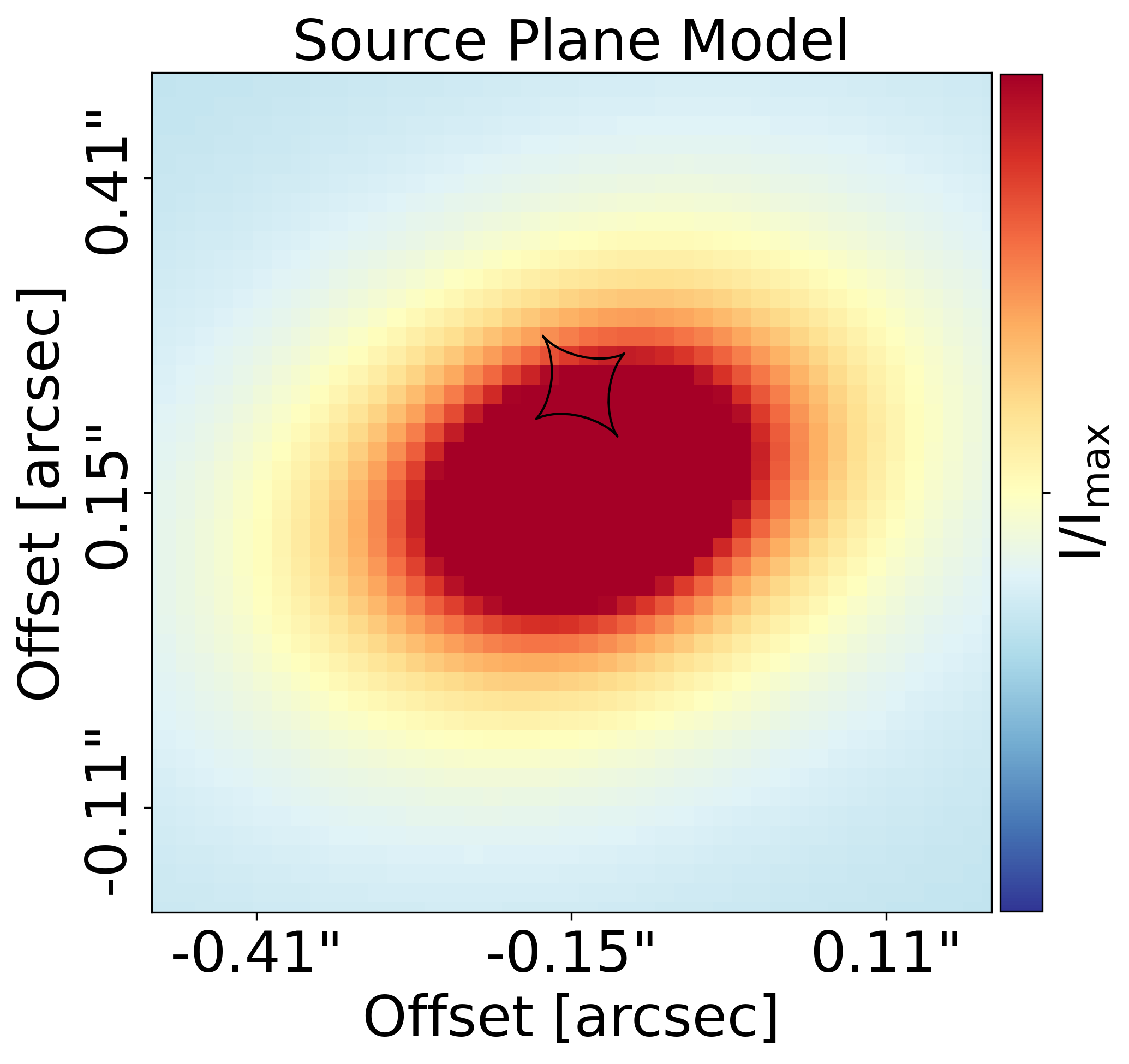}

    \includegraphics[width = 0.19\linewidth]{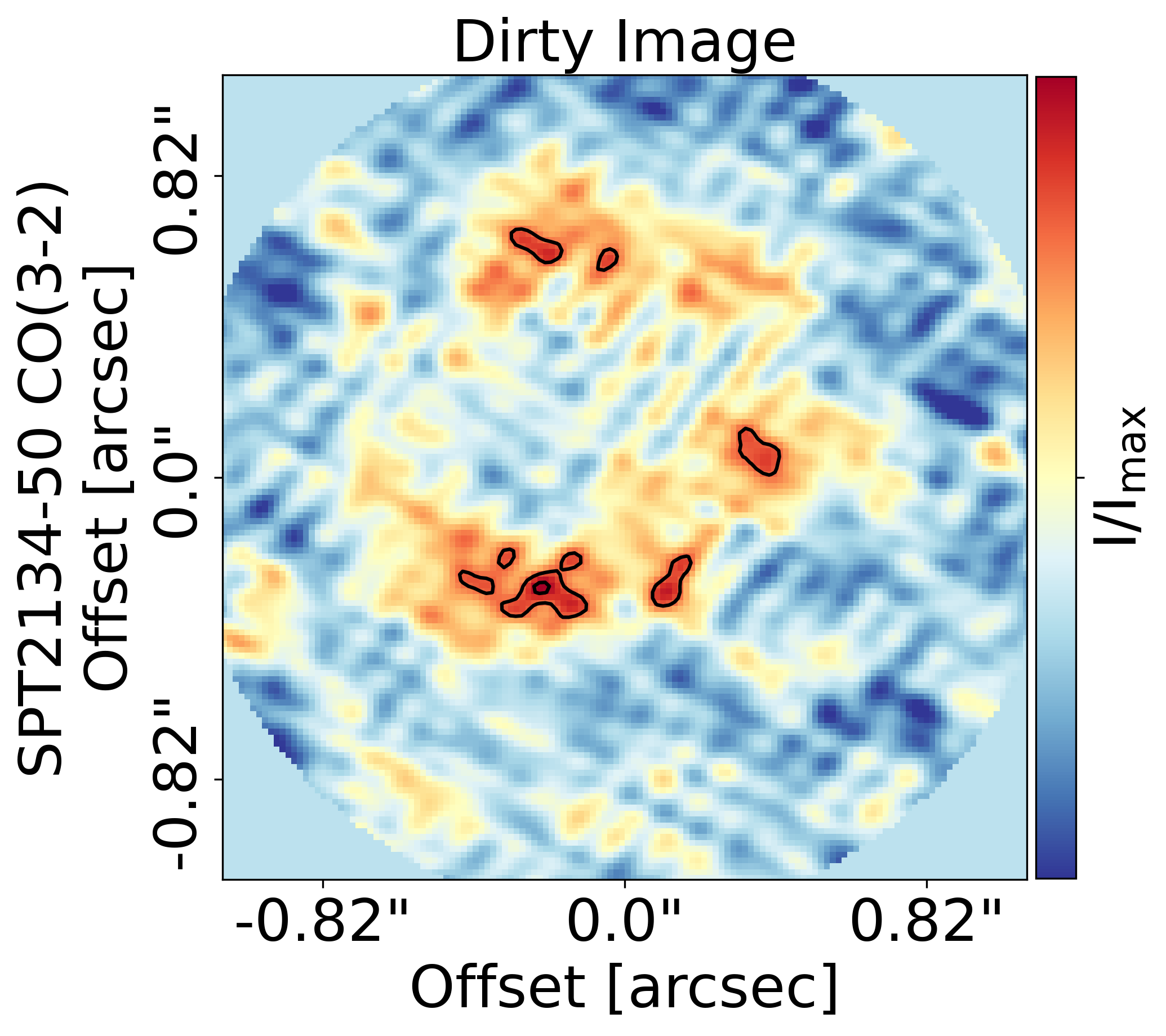}
    \includegraphics[width = 0.18\linewidth]{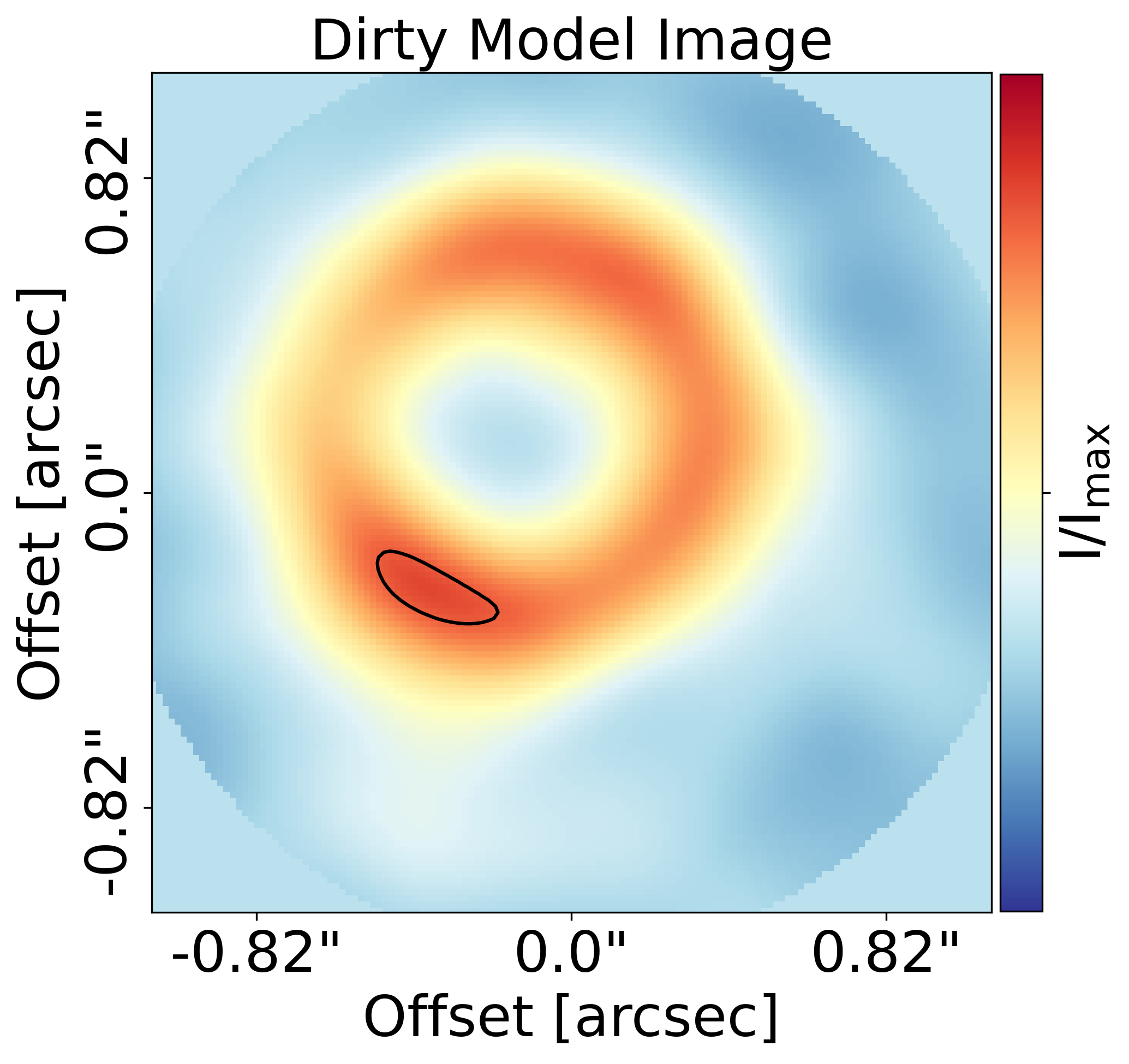}
    \includegraphics[width = 0.18\linewidth]{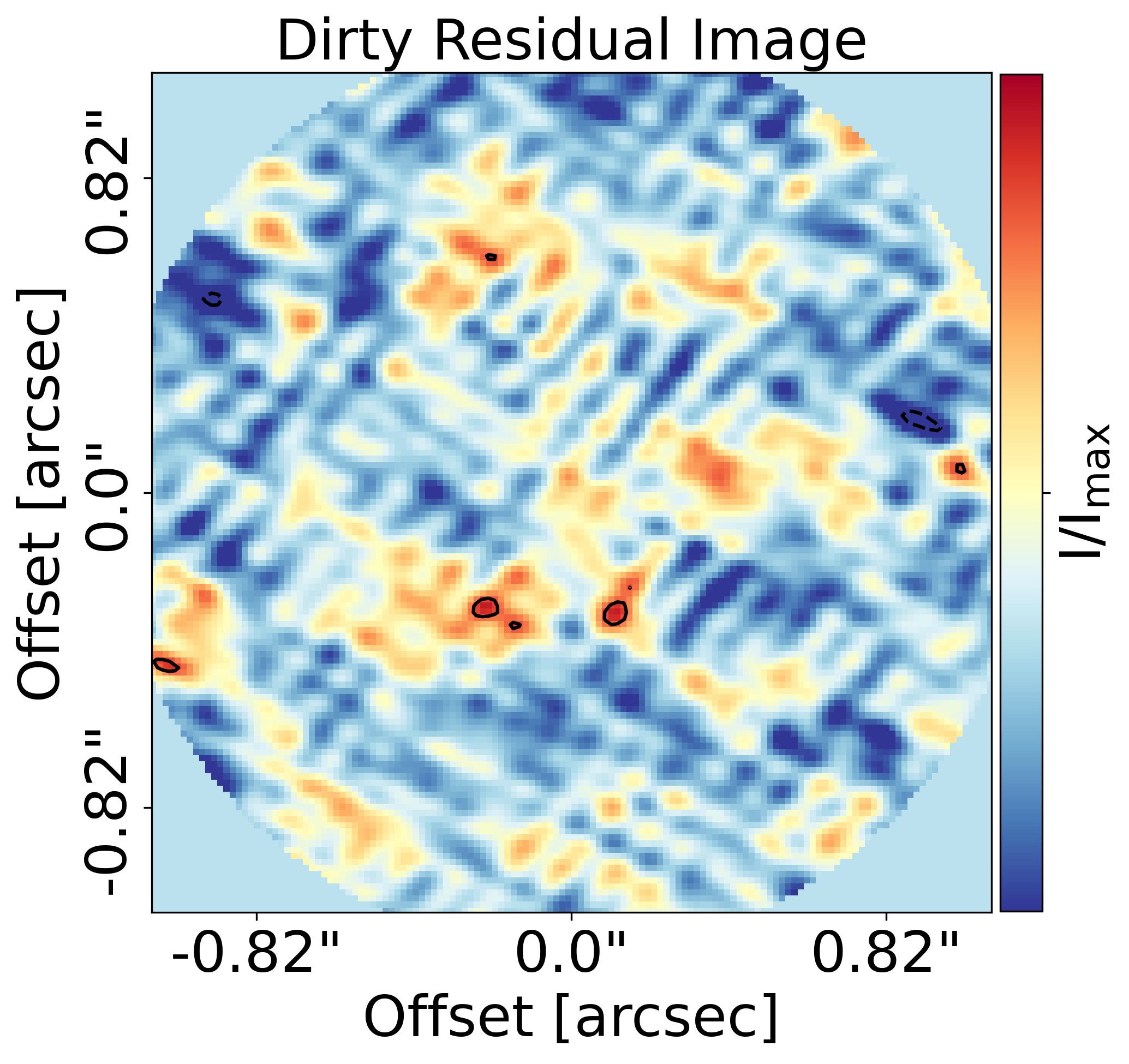}
    \includegraphics[width = 0.18\linewidth]{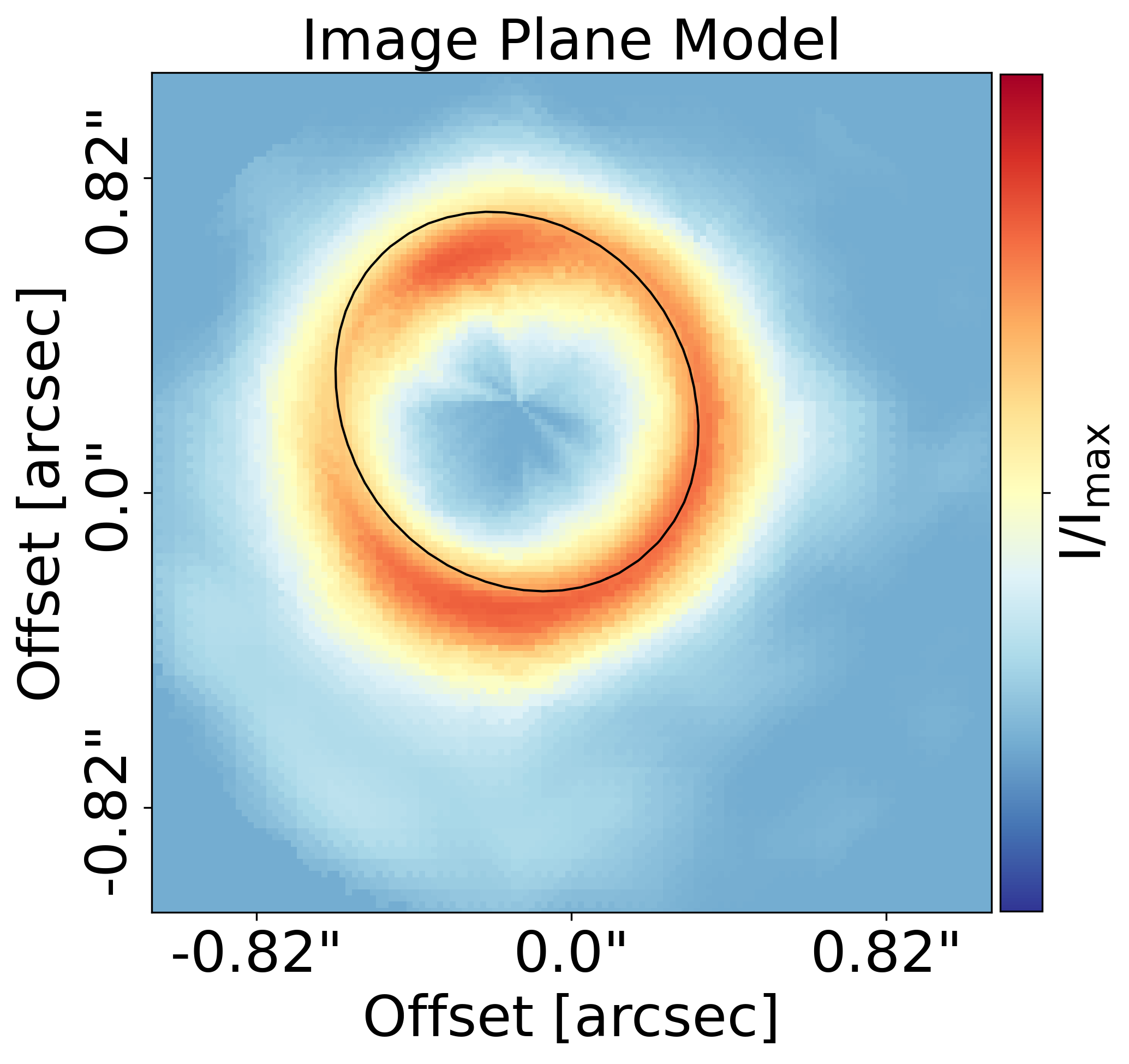}
    \includegraphics[width = 0.18\linewidth]{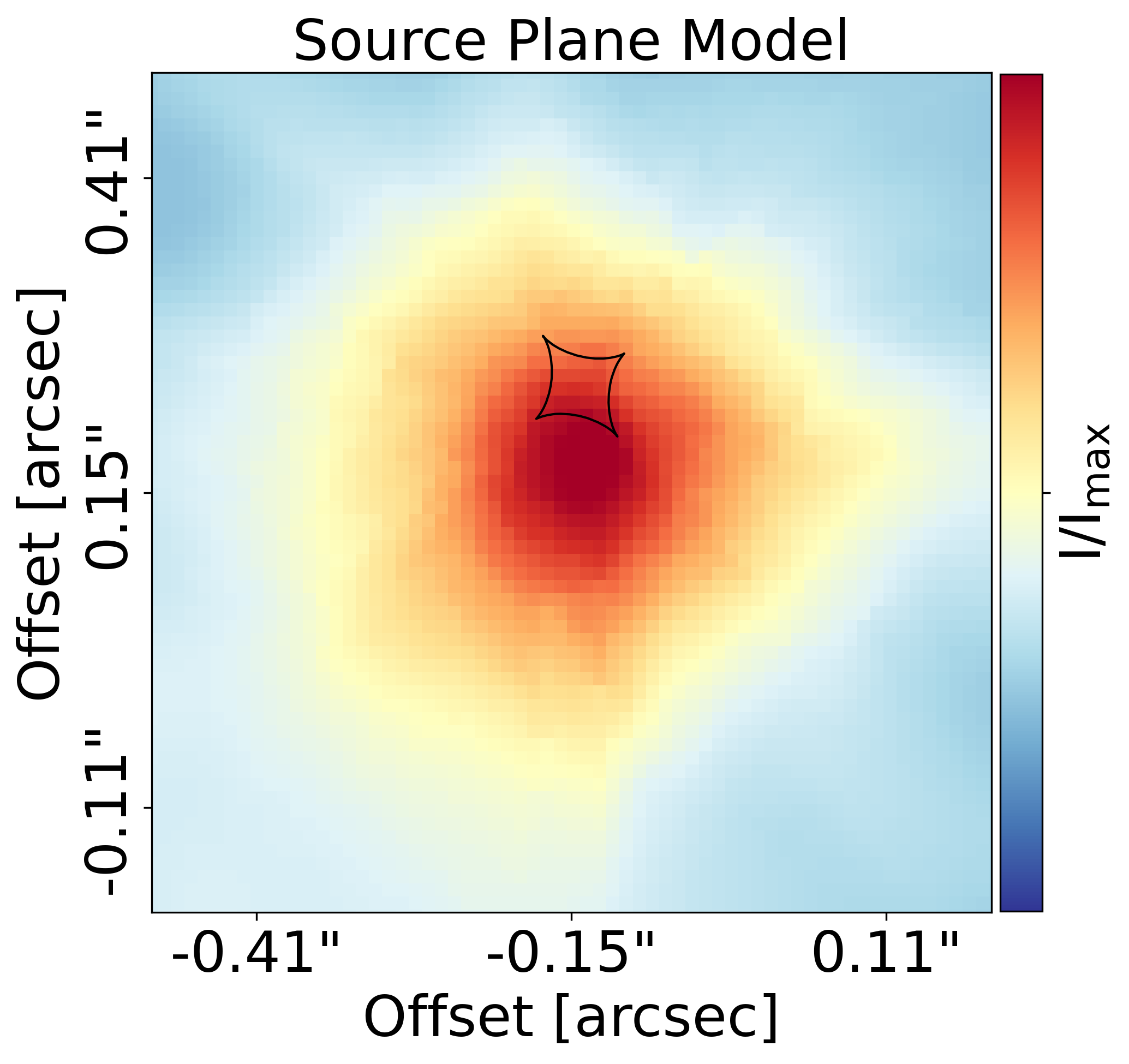}

    \caption{Parametric (rows 1 and 3) and  pixelized (rows 2 and 4) lens modeling of the CO(3--2) emission detected in SPT\,0125-47 (top two rows) and SPT\,2134-50 (bottom two rows). The first column displays the dirty image generated by {\sc PyAutoLens}, with contours at $-3, 3, 4, 5, 6, 7, 8, 9, 10\sigma$ levels. Note: this is not a cleaned image, so structures may differ slightly from those in cleaned images. The second column presents the dirty model image from {\sc PyAutoLens}, also with contours at $-3, 3, 4, 5, 6, 7, 8, 9, 10\sigma$ levels. The third column shows the dirty residual image produced by {\sc PyAutoLens}, with contours at $-3, 3, 4, 5\sigma$ levels. The fourth column illustrates the image plane emission parametric/pixelized model of the data, produced by {\sc PyAutoLens}, with the black line representing the critical line. The fifth column shows the source plane emission parametric+pixelized model of the data from {\sc PyAutoLens}, with the black line indicating the caustic line. All images are centered around the ALMA phase center. }
    \label{fig:parametric_lens_models}
\end{figure*}

The lens modeling for both sources was performed using the sophisticated publicly available lens modeling code {\sc PyAutoLens} \citep{Autolens}. {\sc PyAutoLens} has the ability to perform the lens modeling in the $uv$-plane and therefore on the interferometric visibilities rather than modeling the cleaned images. This is necessary for interferometric data as pixels in interferometric images are not independent of one another; therefore, performing lens modeling directly on the images can bias the results of the lens model. In this work, we performed all lens modeling on the {\sc UVCONTSUB} measurement sets from CASA. {\sc PyAutoLens} additionally has the capability of nonparametrically modeling the source emission, which bypasses the need to assume a specific source-plane morphology (e.g., S\'ersic light profiles). We used {\sc PyAutoLens} to perform both parametric and nonparametric modeling of the CO(3--2) emission toward both sources wherein only the channels containing line emission were used in the modeling. We first performed parametric lensing to optimize the lens mass model then perform nonparametric lens modeling using this mass model. We attempted to model the dust continuum emission using both parametric and nonparametric source plane models, but found that the signal-to-noise (S/N) of the data was too poor to obtain reasonable results. We therefore limited the modeling to the CO(3--2) emission.

\subsubsection{Parametric source modeling} \label{subsubsec:parametric_lens_modeling}

We first performed parametric modeling to establish a working lens model. This is often done on continuum emission as it can be stronger than the corresponding line emission. This method has been used previously for both SPT\,0125-47 and SPT\,2134-50 \citep{Spilker16}. Here, we performed lens modeling directly on the CO(3--2) emission and subsequently applied the best-fit lens model to the continuum data due to the continuum data having significantly lower S/N than the CO(3--2) emission.

Both sources are in galaxy-galaxy lensing morphologies. For both sources we model the lens as a single isothermal ellipsoid (SIE) mass distribution assumed to not have light distributions in the frequencies covered by the ALMA observations. These lens mass distributions were parameterized by their ($x, y$) offset from the observational phase center, Einstein radius, axis ratio, and position angle. We fixed the centers of the mass distributions to be in the center of the observed Einstein rings using uniform priors. We set an upper limit on the Einstein radius for both sources as slightly larger than the Einstein radius reported for the sources in \citet{Spilker16}. All other parameters were free. 

\begin{figure*}
    \centering
    \includegraphics[width = 1.0\linewidth]{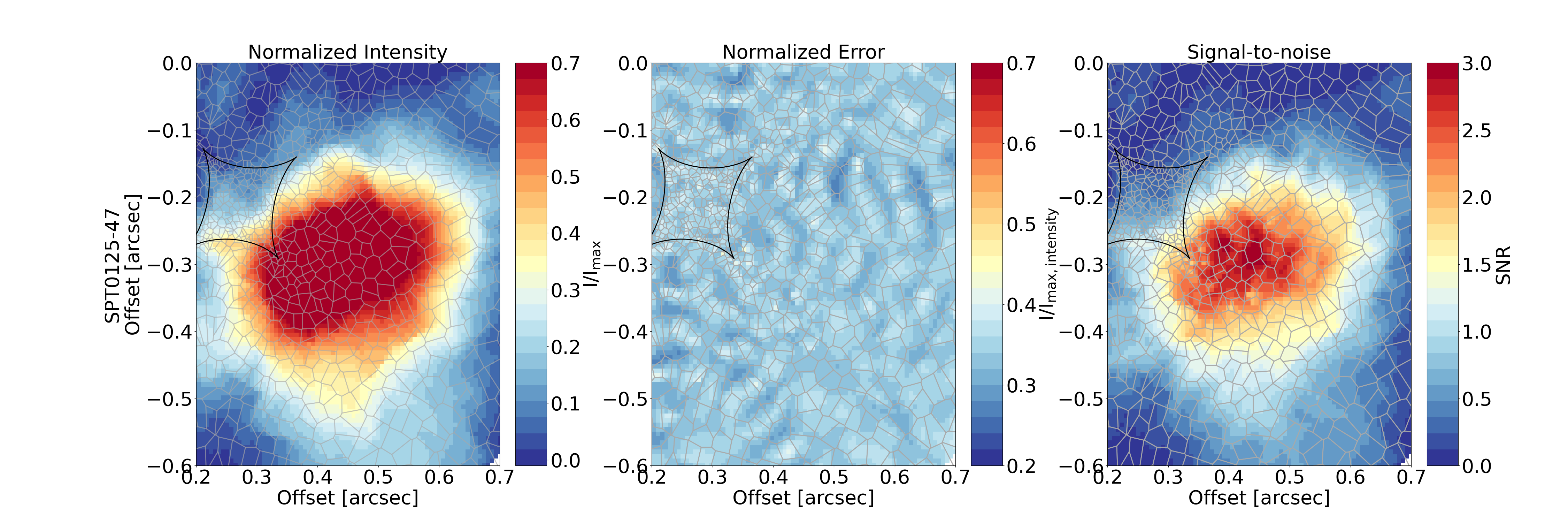}
    \includegraphics[width = 1.0\linewidth]{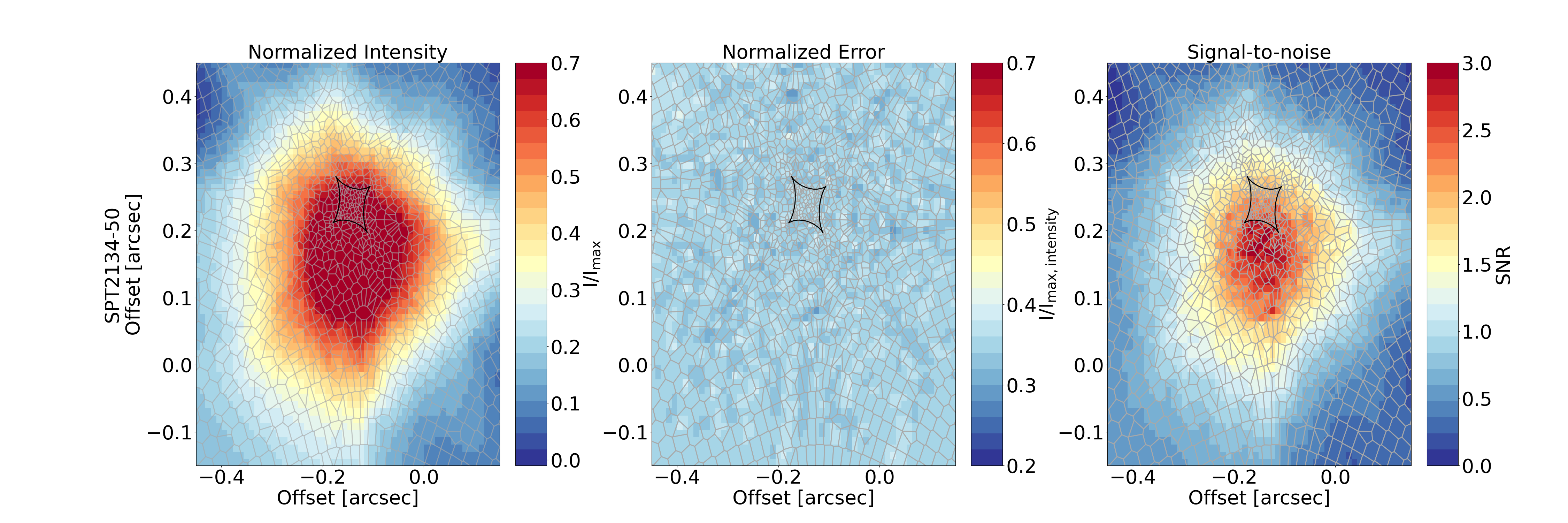}
    \caption{Source plane intensity (left), intensity error maps (middle), and S/N maps (i.e., intensity divided by intensity error, right) for SPT\,0125-47 and SPT\,2134-50. The gray polygons show the Voronoi mesh used for the pixelized reconstruction and the black line shows the caustic line. Both the intensity and intensity error maps are normalized, where the error maps have been normalized to the maximum of the intensity map, meaning that the value can be seen as a percentage error per pixel. The S/N map provides a measure of the significance of the reconstructed emission.}
    \label{fig:error_maps}
\end{figure*}

The sources were parameterized as S\'ersic light distributions using linear light profiles. Linear light profiles minimize the degeneracy between the effective radius and the intensity of the source by solving for both of these parameters using linear algebra. The sources were parameterized by their ($x, y$) offset from the observational phase center, axis ratio, position angle, effective radius, and S\'ersic index. We provide best-fit source and lens parameters in Table \ref{tab:bestfit_lens_model} and Table \ref{tab:lensing_bestfit_sourceparams}. We show the images, models, and residuals for both SPT\,0125-47 and SPT\,2134-50 in Fig. \ref{fig:parametric_lens_models}. 

We note that the primary interest of this paper is the results of nonparametric lens modeling and, therefore, the parametric lens modeling was only performed to: i) optimize the lens mass model, which was then used for the nonparametric fitting; ii) obtain lens magnification factors, and iii) compare our results with literature values.

\subsubsection{Nonparametric source modeling} \label{subsubsec:nonparametric_lens_modeling}

After establishing an optimized lens model from the parametric lens modeling, we used this lens model to nonparametrically model the source emission. This methodology employs an adaptive Voronoi mesh, maintaining regions of higher magnification and, thus, higher angular resolution, around the caustic lines in the source plane. {\sc PyAutoLens} uses a regularization scheme when performing the nonparametric modeling to maintain a balance between over-smoothing and over-fitting the source plane emission \citep{Autolens}. Here, we used a constant regularization scheme, which was parameterized by a regularization coefficient that regulated the smoothness of the emission. The images are then interpolated to a square grid for easier analysis. We used a similar methodology to what has been employed for other studies that perform nonparametric lensing using {\sc PyAutoLens} \citep[e.g.,][]{Maresca22, Giulietti23, Perrotta23, Amvrosiadis25}. We show the images, models, and residuals for both the nonparametric models of the CO(3--2) for both SPT\,0125-47 and SPT\,2134-50 in Fig. \ref{fig:parametric_lens_models}. 

We note that the errors reported in Tables \ref{tab:bestfit_lens_model} and \ref{tab:lensing_bestfit_sourceparams} are formal statistical errors from the sampling algorithms in the lens model fitting. This does not provide a good estimation of the systemic errors, in particular, those associated with the nonparametric lensing. We created source plane error maps with {\sc PyAutoLens} using the same procedure described above, but using the visibility errors, extracted using CASA's STATWT task, instead of the CO(3--2) emission. These error maps show the rms in each Voronoi cell, interpolated onto a square grid. The maps demonstrate how errors propagate from the image plane to the source plane, making it straightforward to produce a S/N map in the source plane. We show the intensity maps, error maps, and S/N maps in Fig. \ref{fig:error_maps}. The error map has been normalized by the maximum of the intensity map and it can therefore be interpreted as a per-pixel percentage error. 

\subsubsection{SPT\,0125-47} 
We found a good parametric lens mass model for the CO(3--2) emission with no residuals at $>3\sigma$ levels. Parametric modeling of this source required using a single-lens mass distribution and, similarly, a single S\'ersic source profile, unlike the three sources required in \citet{Spilker16}. Through this methodology we found a magnification factor of $\mu_{\rm CO(3-2)} = 10.7 \pm 0.002$. This value is lower than what was found for CO(1--0) emission in \citet{Aravena16}; however, we note that the cited study did not directly model the CO(1--0) emission;  rather, these authors calculated the magnification factor from the width of the emission line. This methodology is not as robust as modeling the line emission and therefore we suggest that the magnification factor reported here is a more accurate representation of the magnification affecting the CO molecular gas in SPT\,0125-47. 

We used the optimized lens model from the CO(3--2) emission to create nonparametric models for the CO(3--2) emission. We find that both the parametric and nonparametric source plane models do not require the use of multiple source S\'ersic profiles, as is the case in \citet{Spilker16}. We further discuss this in Section \ref{subsec:morphology}. 

\subsubsection{SPT\,2134-50}
We found a good parametric lens mass model for the CO(3--2) emission with limited residuals at $\sim3\sigma$ levels, but we note that these residuals do not appear to be directly correlated with the Einstein ring of emission from SPT\,2134-50 and might simply be due to noise. Through this methodology, we find a magnification factor of $\mu_{\rm CO(3-2)} = 7.6 \pm 0.002$. This is very similar to the value reported in \citet{Aravena16}; however, similarly to the case of SPT\,0125-47, we suggest that the methodology employed here is more robust; therefore, this value should be considered the accurate magnification value for the CO emission. As with SPT\,0125-47, we used the optimized lens model from the CO(3--2) emission to nonparametrically model the CO(3--2) emission. We found a very similar source plane morphology between the parametric and nonparametric models for the CO(3--2).

\section{Discussion} \label{sec:discussion}

\subsection{Differential lensing} \label{subsec:diff_lensing}

The spectral line profiles of both SPT\,0125-47 and SPT\,2134-50 exhibit skewed profiles wherein red spectral regions appear weaker than blue spectral regions. We investigated whether this is caused by differential lensing across the line profile by dividing each emission line into red and blue spectral bins. We show these regions in relation to the entire spectrum for both sources in Fig. \ref{fig:spec_with_lensing_bins}. We then performed a parametric lens modeling on the red and blue regions of the spectrum for both sources using the best-fit lens model (the same procedure is described in Section \ref{subsec:lens_modeling}). 

Both sources showed significantly different magnification factors between the red and blue regions of the spectrum. For SPT\,0125-47, we found $\mu_{\rm blue} \sim 14$ and $\mu_{\rm red} \sim 9$. For SPT\,2134-50, we found $\mu_{\rm blue} \sim 14$ and $\mu_{\rm red} \sim 4.4$. Full per-channel spectral lensing correction \citep[e.g.,][]{Kade24} is not feasible given the relatively low per-channel S/N of these observations ($\lesssim5\sigma$ per channel). Higher sensitivity data or observations of a brighter emission line could potentially allow for a full reconstruction. These red and blue models provide clear evidence that the skewed spectral profiles and, thus, the necessity of fitting with two Gaussian profiles, result, at least in part, from differential lensing. In this scenario, bluer regions of the spectrum experience higher magnification and therefore would be less bright in a per-channel magnification-corrected spectrum, resulting in a wider and flatter line profile. We note that this could also result in a profile resembling a double-horned profile, which could be associated with rotation; we provide a short discussion on this in Section \ref{subsec:morphology}. We caution that future studies of these two sources should consider the possible effect of differential lensing on, for instance, kinematical modeling.

\begin{figure*}
    \centering
    \includegraphics[width=1.0\linewidth]{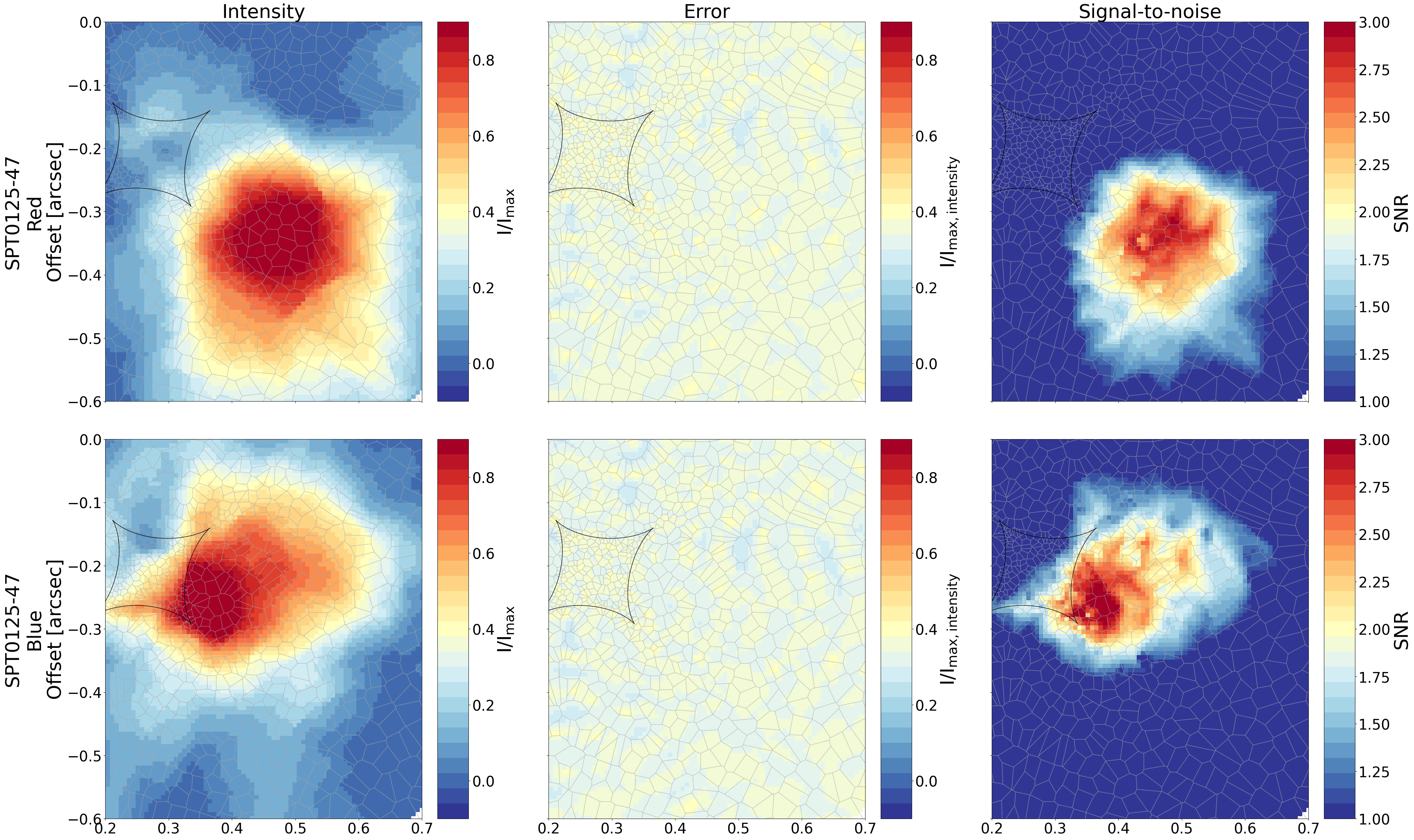}
    \caption{Pixelized source plane models for the red and blue bins for SPT\,0125-47, as described in Section \ref{subsubsec:kinematics} with source plane intensity (left), intensity error maps (middle), and S/N maps (i.e., intensity divided by intensity error, right). The gray polygons show the Voronoi mesh used for the pixelized reconstruction and the black line shows the caustic line. Both the intensity and intensity error maps are normalized, where the error maps have been normalized to the maximum of the intensity map, meaning that the value can be seen as a percentage error per pixel.}
    \label{fig:redblue_bins_spt0125}
\end{figure*}

\subsection{Source plane properties} \label{subsec:morphology}

Recent studies have demonstrated the importance of nonparametric, pixelized, source plane models of lensed galaxies to accurately interpret lensed source properties including kinematics and morphology \citep[e.g.,][]{Rybak20, Rizzo21, Giulietti23, Perrotta23, Amvrosiadis25}. Although the approximate morphology of the parametric and nonparametric modeling are consistent with each other with regard to the location of the background source (e.g., in relation to the caustic line), the main discrepancy lies in the lack of detail in the parametric models. This highlights the importance of modeling that avoids making a priori assumptions about the source plane morphology.  We discuss the morphology and kinematics of the two sources in the pixelized source plane models below. 

\subsubsection{Morphology}

SPT\,0125-47 was  previously reported as a system composed of three different galaxies with one member dominating in brightness \citep{Spilker16}. This conclusion was based on parametric lens modeling of the morphology of the $870\,\mu\mathrm{m}$ dust continuum emission. In this study, we found that our parametric lens model does not require multiple sources to obtain a good fit with no significant residual emission. We also found a similar source plane morphology in the nonparametric model.

SPT\,2134-50 was  previously reported as a single galaxy system \citep{Spilker16}, which is in good agreement with the results of the parametric lens modeling in this work. In general, we found that both SPT\,0125-47 and SPT\,2134-50 have smooth source plane morphologies in the nonparametric models.

We performed a brief investigation into whether we should expect to be able to resolve clumps in source plane models based on the expected size of clumps found in previous studies. For SPT\,0125-47, assuming a static magnification factor across the image, the source-plane beam is $0''.033 \times 0''.033$, corresponding to $\sim270$\,pc. This means that we can only clearly resolve features on scales larger than $\sim800$\,pc. For SPT\,2134-50, under the same assumption of the static source plane beam, the source-plane beam is $0''.054 \times 0''.047$, corresponding to $\sim400$\,pc. This means that we can only clearly resolve features on scales larger than $\sim1''.2$. Values for the expected size of clumps in the interstellar medium (ISM) as found by simulations and observations are significantly smaller than these scales \citep[e.g.,][]{Dekel09, Bournaud14, Romeo14, Iono16, Oteo17}. For example, \citet{Spilker22} studied the $z =6.9$ DSFG SPT\,0311–58 and found clumps on scales of a few hundred parsecs; in addition, we note that similar scales were found in \citet{Rybak20}. Therefore, even at the relatively high angular resolutions of the observations in this work, it is not possible to resolve features on the scale of the expected  clump size, regardless of any of the lens modeling uncertainties or difficulties.

\subsubsection{Kinematics}\label{subsubsec:kinematics}

Studies such as \citet{Rizzo21} and \citet{Amvrosiadis25} have performed kinematical analyses on lensed galaxies, in particular, lensed SPT sources. This type of analysis requires the creation of a de-lensed source plane emission cube. The relatively low per-channel S/N of the CO(3--2) emission prevents the construction of a fully delensed source plane spectral cube. Source plane models based on the entire integrated emission line already exhibit substantial uncertainties. Only a few regions achieve high significance (S/N $>5$), while most regions are at S/N $\sim 2-3$ (see Fig. \ref{fig:error_maps}). We performed two initial attempts to construct a delensed source plane using a restricted number of channels centered on the brightest regions of the spectrum, selecting approximately ten channels per source. However, the S/N in the individual channels was insufficient to enable reliable source-plane modeling. A further attempt using only three channels also yielded unsuccessful results.

\begin{figure*}
    \centering
    \includegraphics[width=1.0\linewidth]{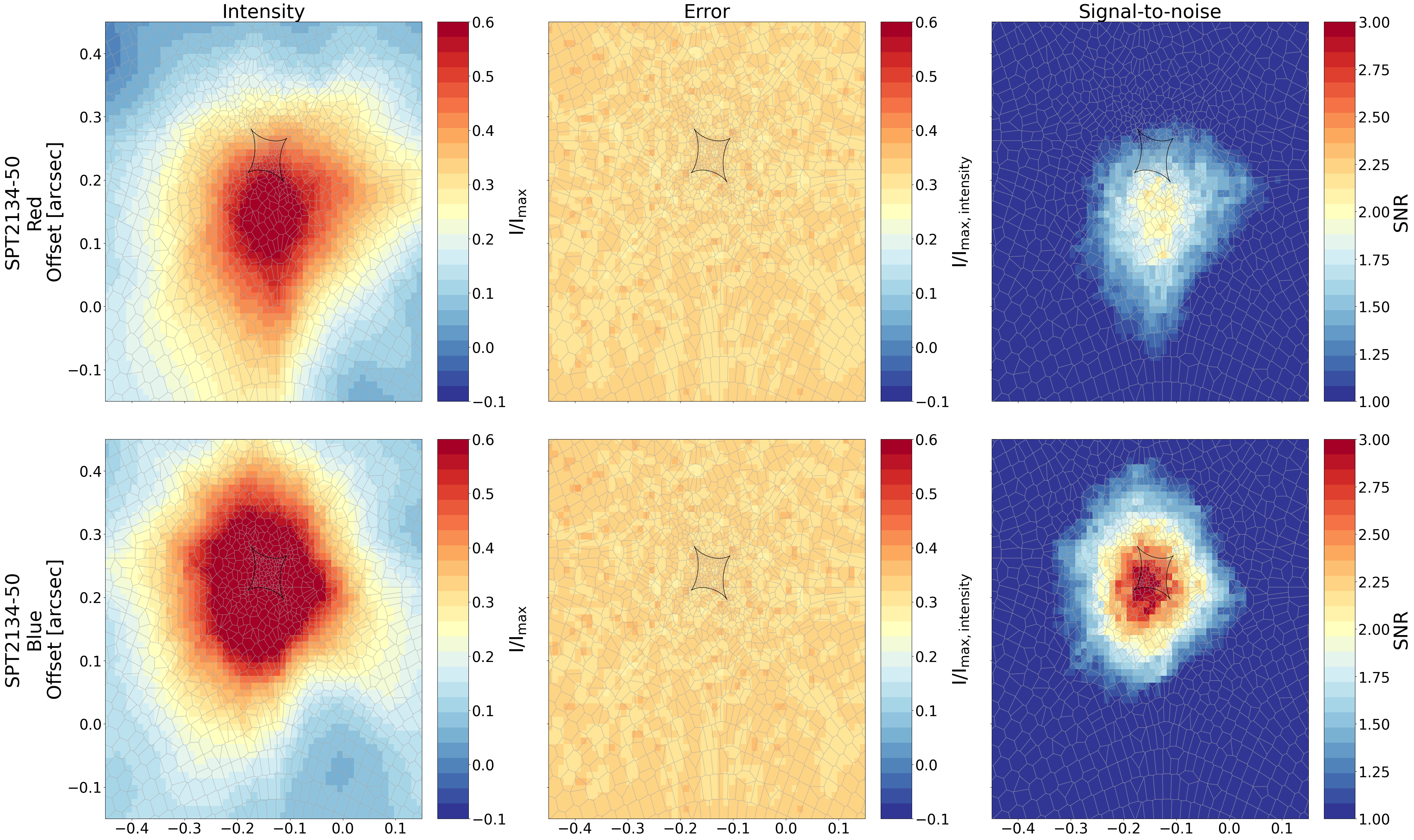}
    \caption{Pixelized source plane models for the red and blue bins for SPT\,2134-50 as described in Section \ref{subsubsec:kinematics} with source plane intensity (left), intensity error maps (middle), and S/N maps (i.e., intensity divided by intensity error: right). The gray polygons show the Voronoi mesh used for the pixelized reconstruction and the black line shows the caustic line. Both the intensity and intensity error maps are normalized, where the error maps have been normalized to the maximum of the intensity map, meaning that the value can be seen as a percentage error per pixel.}
    \label{fig:redblue_bins_spt2134}
\end{figure*}

Instead, we used the red and blue spectral bins (Section \ref{subsec:diff_lensing}) to create pixelized source-plane models for these bins for each source. While this approach yielded acceptable results, the S/N remains low with considerable source plane uncertainties. For example, in the case of SPT\,2134-50, there are no regions of the image with S/N $>3$ in the red bin. The channel maps and associated uncertainty and S/N maps are shown in Fig. \ref{fig:redblue_bins_spt0125} and \ref{fig:redblue_bins_spt2134}. We found tentative evidence of a velocity gradient across both sources, which might be indicative of more ordered rotation in these sources, although the current data do not allow for the construction of reliable source-plane moment-1 maps, which would be required to further investigate this possibility. Given the quality of the reconstructions, we urge caution when drawing conclusions based on these source plane reconstructions. Deep, high-sensitivity observations of a bright emission line such as \cii or higher-$J$ CO lines, with comparable or higher angular resolution, would be necessary to confirm and characterize these possible velocity structures. 

\subsubsection{Implications}

We calculated the depletion time ($\rm t_{dep} = SFR/M_{gas}$) for both sources using the total molecular gas mass and IR luminosity from \citet{Aravena16} and following the procedure within. This value is not dependent on the lensing magnification. We found a depletion time of 0.051\,Gyr for SPT\,0125-47 and 0.037\,Gyr for SPT\,2134-50. These values are in good agreement with the depletion times found for other SPT sources \citep{Aravena16}. Both sources lie well below the expected depletion timescales for main sequence (MS) galaxies from \citet{Saintonge13} as expected given their significantly elevated SFRs above MS galaxies. The high SFRs and short depletion times could be interpreted as an indication that both systems have recently experienced an interaction and/or merger, which could explain their high IR luminosities and how these sources had obtained sufficient gas to sustain such high SFRs. 

This could represent a situation similar to the DSFG G09v1.97 at $z = 3.63$ \citep{Kade_sub}. This is an attractive explanation as it follows the classical picture of massive galaxy evolution \citep[e.g.,][]{Hopkins08}, but it is only one of many possibilities which, given the current data quality, are infeasible to verify. For example, it is entirely conceivable that improved observations of a stronger emission line would show these sources breaking into multiple sources, possibly as part of the process of merging or simply by virtue of being spatially close to each other. An additional scenario is that these sources represent examples of more secular evolution, wherein cold mode accretion is the primary galaxy growth mechanism in operation \citep[e.g.,][]{keres05}.

Our analysis demonstrates that high-S/N data are essential for reliable pixelized source plane reconstructions. Given the observational constraints of these data, these results should be considered as preliminary and require confirmation through deeper observations or observations of brighter emission lines (e.g., [C\,{\sc ii}]). Although pixelized reconstructions remain the optimal method for interpreting strongly gravitationally lensed galaxies, future benchmarking studies are necessary to establish the S/N requirements for robust kinematic analyses. However, such work falls outside the scope of this paper. These findings underscore the value of pixelized source plane reconstructions, and the caution necessary when interpreting results from this method, while highlighting the observational difficulties inherent to high-redshift ISM studies.

\section{Conclusions} \label{sec:conclusions}
In this paper, we  present the detection of high-resolution ALMA observations of CO(3--2) emission in two lensed sources, SPT\,0125-47 and SPT\,2134-50. We describe the  parametric and nonparametric lens modeling we performed on both galaxies using {\sc PyAutoLens}. Our conclusions are listed below.

\begin{enumerate}
    \item We have further improved the lens models of SPT\,0125-47 and SPT\,2134-50 using the publicly available code {\sc PyAutoLens}. We find that both sources require only a single S\'ersic profile as a description of the background source to obtain a good lens model, in contrast to previous findings. 

    \item We found clear evidence of differential lensing across the spectrum for both SPT\,0125-47 and SPT\,2134-50. With the current data availability, a full per-channel magnification correction is not feasible. However, we emphasize that the effect is not insignificant and should be taken into account in future studies of these two sources. 

    \item We divided the line profile of both sources into two red and blue bins of the spectra and performed pixelized reconstructions of these bins.  In the two-bin scenario, we found evidence of a velocity gradient, but given the very tentative nature of this detection, we did not investigate this possibility further.

    \item We find that the parametric and pixelized models of both SPT\,0125-47 and SPT\,2134-50 suggest a single background source. However, we note that the source plane resolution is not sufficient to conclusively determine whether both of them are composed of multiple smaller galaxies in an ongoing merger or a single disk.

\end{enumerate}

Given the very high SFRs and low depletion times found in previous studies and reported here, combined with the lack of evidence of interactions or mergers in our source plane emission models, we suggest that these sources have recently undergone an interaction or merger that triggered the high SFR. In such a case, these sources would both be in the process of settling into disks. Very-high-angular-resolution observations of a brighter emission line, such as [C\,{\sc ii}], combined with pixelized source plane modeling would be necessary to arrive at robust conclusions on the true nature of these two sources.

\begin{acknowledgements}

We thank the anonymous referee for constructive feedback that helped improve the clarity of the manuscript. K.K. acknowledges support from the Nordic ALMA Regional Centre (ARC) node based at Onsala Space Observatory. The Nordic ARC node is funded through Swedish Research Council grant No. 2017-00648. K.Kn. acknowledges support from the Knut and Alice Wallenberg Foundation (KAW 2017.0292). S.K. gratefully acknowledges funding from the European Research Council (ERC) under the European Union’s Horizon 2020 research and innovation programme (grant agreement No. 789410). 

This paper makes use of the following ALMA data: ADS/JAO.ALMA\#2016.1.01231.S. ALMA is a partnership of ESO (representing its member states), NSF (USA) and NINS (Japan), together with NRC (Canada), NSTC and ASIAA (Taiwan), and KASI (Republic of Korea), in cooperation with the Republic of Chile. The Joint ALMA Observatory is operated by ESO, AUI/NRAO and NAOJ.

\end{acknowledgements}

\bibliographystyle{aa} 
\bibliography{biblio}

\appendix

\section{Best-fit parametric modeling parameters}
Best-fit parametric lens and source models for SPT\,0125-47 and SPT\,2134-50.

\begin{table*}[]
    \centering
    \caption{Best-fit parametric lens models for both SPT\,0125-47 and SPT\,2134-50.}
    \begin{tabular}{l c c c c c c} \hline \hline
        Lens & $z^{a}$ & $x_{\mathrm{off}}^{b}$ & $y_{\mathrm{off}}^{c}$ & q$^{d}$ & PA$^{e}$ & Einstein radius \\
         & & [''] & [''] & & [degrees] & [''] \\ \hline
        \vspace{2.0mm}
        SPT\,0125-47$_{\rm lens}$ & 0.31 & $0.18^{+0.16}_{-0.22}$ & $-0.16^{+0.13}_{-0.08}$ & $0.71^{+0.44}_{-0.21}$ & $48.9^{+27.5}_{-27.8}$ & $0.96^{+0.16}_{-0.06}$ \\

        SPT\,2134-50$_{\rm lens}$ & 0.78 & $-0.15^{+1.0}_{-0.76}$ & $0.24^{+0.66}_{-1.1}$ &  $0.71 ^{+0.57} _{-0.28}$ & $-31.19 ^{+58.50} _{-120.82}$ & $0.48^{+0.19}_{-0.46}$ \\ \hline 
    \end{tabular}
    
    \tablefoot{
        \tablefoottext{a}{Redshift of the lens.}
        \tablefoottext{b}{x-position of the lens defined in offset from the observational phase center.}
        \tablefoottext{c}{y-position of the lens defined in offset from the observational phase center.}
        \tablefoottext{d}{Minor-to-major axis ratio of the lens.}
        \tablefoottext{e}{Position angle (PA) of the lens, defined by {\sc PyAutoLens} as counter-clockwise from the x-axis.}}
    \label{tab:bestfit_lens_model}
\end{table*}

\begin{table*}[]
    \centering
    \caption{Best fit parametric source models for both SPT\,0125-47 and SPT\,2134-50.}
    \begin{tabular}{l c c c c c c c} \hline \hline
       Source & $x_{\mathrm{off}}^{a}$ & ${y_\mathrm{off}}^{b}$ & q$^{c}$ & PA$^{d}$ & Effective radius & S\'ersic index & $\mu^{e}$ \\
         & [''] & [''] & & [degrees] & [''] & \\ \hline

        SPT\,0125-47 & $0.41 \pm 0.07$ & $-0.27^{+0.10}_{-0.03}$ & $0.63 ^{+0.36} _{-0.35}$ & $60^{+150} _{-30}$ & $0.28^{+6.2}_{-0.17}$ & $1.5^{+3.5}_{-0.71}$  & 10.7\\

        SPT\,2134-50 & $-0.14^{+0.26}_{-0.29}$ & $ 0.15^{+0.13}_{-0.51}$ & $0.58 ^{+0.44} _{-0.40}$ & $22 ^{+111} _{-68}$ & $1.45^{+27.4}_{-1.4}$ & $4.0^{+0.98}_{-3.19}$ & 7.64 \\
     \hline 
    \end{tabular}
    
    \tablefoot{
        \tablefoottext{a}{x-position of the source defined in offset from the observational phase center.}
        \tablefoottext{b}{x-position of the source defined in offset from the observational phase center.}
        \tablefoottext{c}{Minor-to-major axis ratio of the source.}
        \tablefoottext{d}{Position angle of the source, defined by {\sc PyAutoLens} as counter-clockwise from the x-axis.}
        \tablefoottext{e}{Magnification factor of the emission calculated through the ratio of the image plane to source plane emission.}}
    \label{tab:lensing_bestfit_sourceparams}
\end{table*}

\section{Spectra with red and blue bins}
Spectra of both sources showing the red and blue lensing bins used to investigate the differential lensing and the results of the parametric lens modeling for both sources.

\begin{figure*}
    \centering
    \includegraphics[width = 0.45\linewidth]{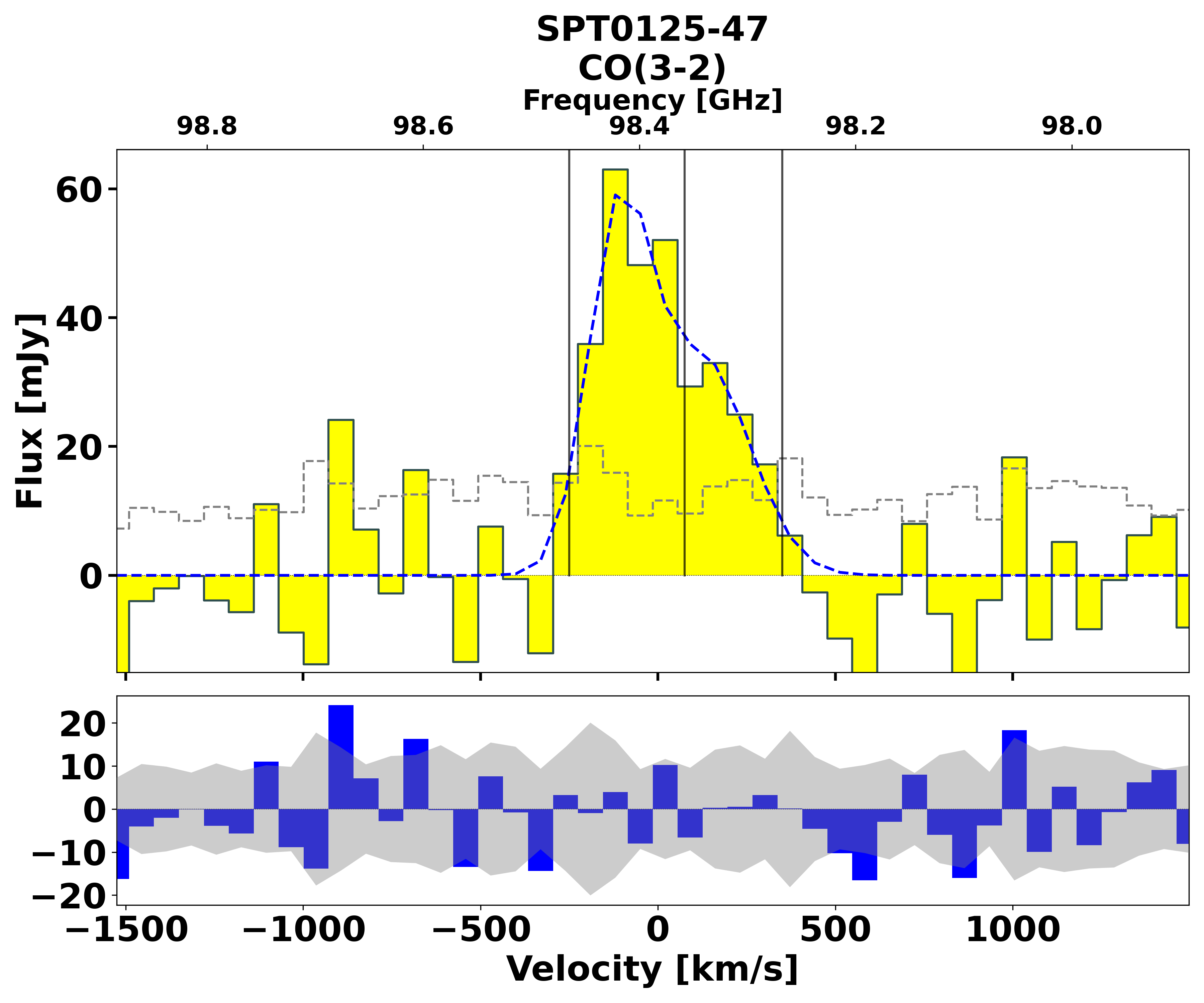}
    \includegraphics[width = 0.45\linewidth]{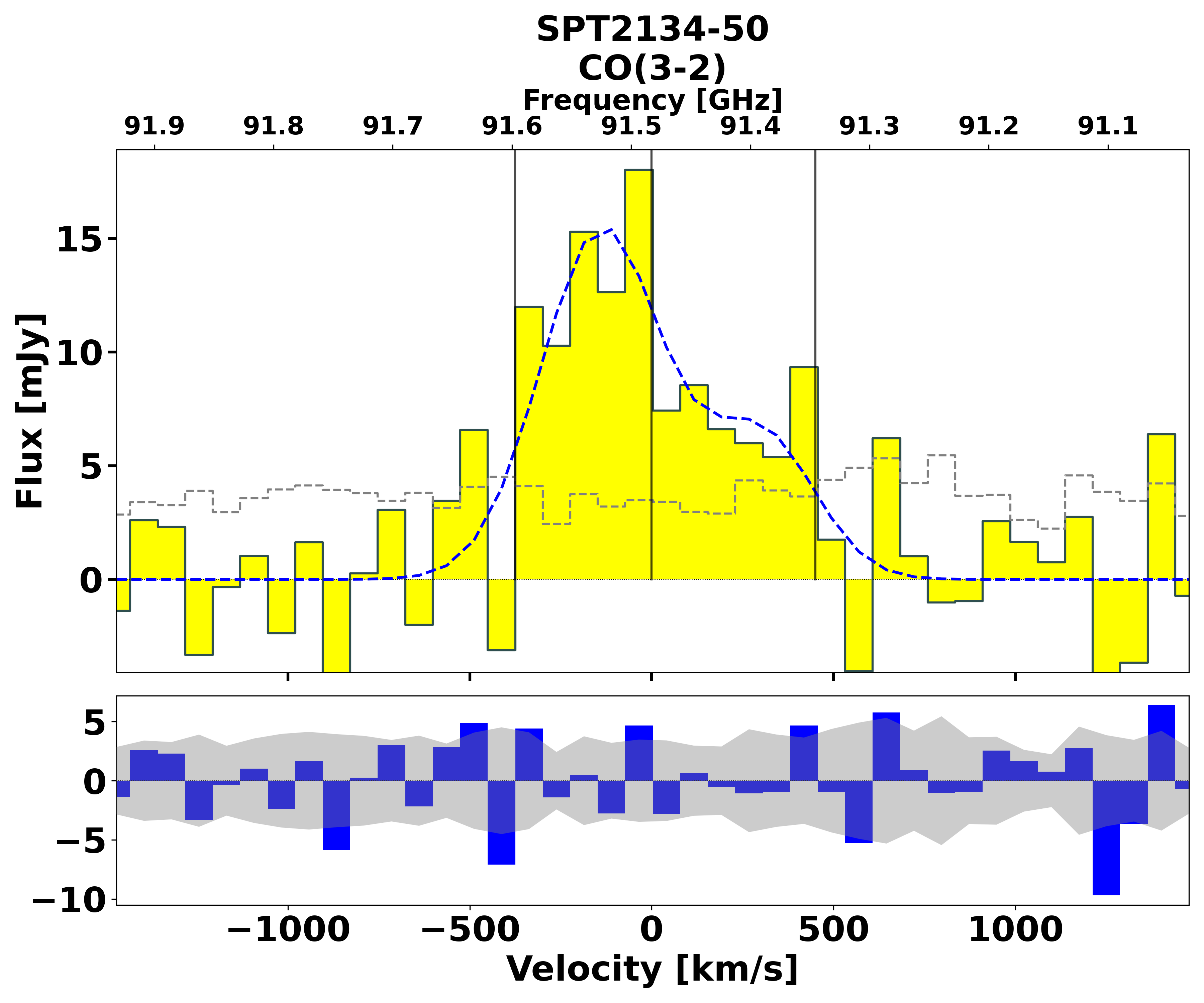}
    \caption{Spectra of the CO(3--2) toward SPT\,0125-47 (left) and SPT\,2134-50 (right) showing the coverage of the red and blue bins used to investigate the effect of differential lensing, as described in Section \ref{subsec:diff_lensing}. In both cases, the spectrum is shown in the top panel, and the residuals from the Gaussian fit are shown in the bottom panel. The dashed blue line shows the two Gaussian fit to the spectra. The dashed gray line in the top panel and the shaded gray region in the lower panel indicate the per-channel RMS. Additionally, the top axis in the top panel of each spectrum displays the corresponding frequency.}
    \label{fig:spec_with_lensing_bins}
\end{figure*}

\end{document}